\documentclass{interact}

\newif\ifblind

\usepackage{amsmath,amssymb,amsthm,amsfonts,bm,graphics,graphicx,url,multicol,multirow,longtable,lscape,float,graphicx,booktabs,setspace,natbib}

\usepackage{lineno}

\usepackage{xr-hyper}
\usepackage{hyperref}
\hypersetup{
    colorlinks=true,
    linkcolor=blue,
    filecolor=black,      
    urlcolor=black,
    citecolor=blue
}
\usepackage[capitalise,noabbrev]{cleveref}

\newcommand{\bbeta}{{\bm \beta}}

\newcommand{\bS}{\mathbf{S}}

\newcommand{\bSigma}{\bm{\Sigma}}

\newcommand{\bepsilon}{\bm{\epsilon}}

\newcommand{\bC}{\mathbf{C}}
\newcommand{\bc}{\bm{c}}

\newcommand{\diag}{\mathrm{diag}}
\newcommand{\tr}{\mathrm{tr}}
\newcommand{\bX}{\mathbf{X}}

\newcommand{\bY}{\mathbf{Y}}
\newcommand{\by}{\mathbf{y}}

\newcommand{\bx}{\bm{x}}

\newcommand{\bz}{\bm{z}}

\newcommand{\bTheta}{\bm{\Theta}}

\newcommand{\hatbSigma}{\widehat{\bm{\Sigma}}}
\newcommand{\bPsi}{\bm{\Psi}}
\newcommand{\hatbPsi}{\widehat{\bPsi}}
\newcommand{\bpi}{\bm{\pi}}
\newcommand{\bpsi}{\bm{\psi}}

\newcommand{\hatbbeta}{\widehat{\bbeta}}

\newcommand{\bZ}{\mathbf{Z}}

\newcommand{\bs}{\bm{s}}

\newcommand{\cmark}{\ding{51}}%
\newcommand{\xmark}{\ding{55}}%

\def\E{{\rm E}\,}

\usepackage{amsmath}
\usepackage{amsfonts}
\usepackage{amssymb}
\usepackage{amsthm}
\usepackage{subcaption}

\usepackage{enumitem}

\usepackage{fullpage}

\usepackage{algorithm}
\usepackage{algpseudocode}

\def\boxit#1{\vbox{\hrule\hbox{\vrule\kern6pt  \vbox{\kern6pt#1\kern6pt}\kern6pt\vrule}\hrule}}

\usepackage{pifont}

\theoremstyle{plain}

\theoremstyle{definition}

\theoremstyle{remark}

\begin{document}
\title{Mixture of regressions with multivariate responses for discovering subtypes in Alzheimer's biomarkers with detection limits}

\date{}

\ifblind
\author{Blinded}
\else
\author{
\name{Ganzhong Tian\textsuperscript{a}, John Hanfelt\textsuperscript{a}, James Lah\textsuperscript{b}, Benjamin B. Risk\textsuperscript{a}\thanks{CONTACT B.~B. Risk. Email: brisk@emory.edu}}
\affil{\textsuperscript{a}{Department of Biostatistics and Bioinformatics, Emory University}}
\affil{\textsuperscript{b}{Department of Neurology, Emory University School of Medicine}}
}
\fi

\maketitle


\begin{abstract}
There is no gold standard for the diagnosis of Alzheimer's disease (AD), except from autopsies. Unsupervised learning can provide insight into the pathophysiology of AD. A mixture of regressions can simultaneously identify clusters from multiple biomarkers while accounting for within-cluster demographic effects. Cerebrospinal fluid (CSF) biomarkers for AD have detection limits, which create additional challenges. We apply a mixture of regressions with a multivariate truncated Gaussian distribution (also called a censored multivariate Gaussian mixture of regressions or a mixture of multivariate tobit regressions) to over $3,\!000$ participants from the Emory Goizueta Alzheimer's Disease Research Center and Emory Healthy Brain Study to examine amyloid-beta peptide 1-42 (Abeta42), total tau protein and phosphorylated tau protein in CSF with known detection limits. We address three gaps in the literature on mixture of regressions with a truncated multivariate Gaussian distribution: software availability; inference; and clustering accuracy. We discovered three clusters that tend to align with an AD group, a normal control profile and non-AD pathology. The CSF profiles differed by race, gender and the genetic marker ApoE4, highlighting the importance of considering demographic factors in unsupervised learning with detection limits. Notably, African American participants in the AD-like group had significantly lower tau burden.
\end{abstract}


\doublespacing

\section{Introduction}
\label{s:intro}
A definitive diagnosis of Alzheimer's disease (AD) is only possible from an examination of brain tissue in an autopsy \citep{Dubios2007b}. The problem is made worse by the fact that clinical diagnosis using biomarkers have historically been based on studies dominated by people of European ancestry \citep{Blennow2015}. African American individuals are greatly underrepresented in AD biomarker studies and clinical trials \citep{shin2016underrepresentation}, and CSF biomarker levels differ by race \citep{garrett2019racial}. Unsupervised learning was applied to CSF biomarkers to reveal insights into AD, but race and other demographic factors were not considered \citep{Meyer2010}. There are at least three challenges to analyzing CSF AD biomarker data: 1) multivariate biomarkers have detection limits, resulting in censoring; 2) the disease status is unknown since there is no gold standard; and 3) demographic effects may depend on unknown subtypes. Current statistical software do not simultaneously address these problems (\cref{soft_list}). Our goals are twofold: 1) cluster participants into groups using an unsupervised multivariate method, since no gold standard is available and current criteria may be limited by factors such as European ancestry, and 2) gain insights into pathophysiology by estimating within-cluster effects of demographic variables (race, gender, the genetic marker ApoE4, age and education).

Our study is motivated by the Emory Goizueta Alzheimer's Disease Research Center and the Emory Healthy Brain Study (hereafter, Emory ADRC/HBS Dataset), which contains three CSF biomarkers (amyloid-beta peptide 1-42 [Abeta42], total tau protein [tTau] and phosphorylated tau protein [pTau]) from lumbar punctures of over $3,\!000$ individuals \citep{Goetz2019}. The dataset contains 16.5\% (495) African American participants, which is substantially higher than the Alzheimer's Disease Neuroimaging Initiative ($<5\%$). An important limitation of the assay is that approximately $15\%$ of the participants in the Emory ADRC/HBS dataset have one of the three biomarker levels defined by the detection limits of the assay. In this paper, we define a censored response variable in the same way as other mixture modeling papers \citep{Jedidi1993,Lee2012}: censoring occurs if the value of the response variable is set equal to the detection limit when the true value is more extreme than the threshold, while the predictors are available for all observations (e.g., participants with Abeta42 over $1,\!700$ have their Abeta42 values set equal to $1,\!700$). This differs from truncation, which typically refers to a restricted sampling of the distribution of the population (e.g., if patients with Abeta42 over $1,\!700$ were not recorded, then the data would be truncated).

Unsupervised learning, such as Gaussian mixture models (GMMs), are popular tools for defining disease subtypes when a gold standard is not available \citep{Collins2014}. Model-based clustering approaches derived from GMMs have advantages over distance-based clustering algorithms such as K-means. GMMs estimate posterior probabilities of group membership for each data point rather than hard clustering. GMMs utilize a statistical model that can account for correlations.  From a probabilistic perspective, K-means assumes spherically shaped clusters, which can lead to poor results when features are correlated \citep{Coates2012}. In our application, CSF biomarkers of AD are highly correlated. A Gaussian mixture of regressions (GMR) model, also called ``switching regressions'' in econometrics, extends GMMs to datasets with predictors by modeling the mean structure of each group using regression \citep{Goldfeld1973,Quandt1978}. These models allow for the effect of predictors to be modified by the latent groups. These models have also been extensively discussed in the machine learning literature, where they are called ``mixture of experts" models \citep{Yuksel2012}.  

Censored multivariate Gaussian mixtures of regressions (censored GMRs), also known as a mixture of regressions with a truncated multivariate Gaussian distribution or a multivariate mixture of tobit regressions, have been previously considered in the literature. \cite{Lee2012} derived EM algorithms for fitting multivariate GMMs to censored data. To model predictors, \cite{Jedidi1993} derived an EM algorithm for a mixture of tobit regressions with a univariate censored response, and more recent extensions of tobit regression for a univariate response model errors using a finite mixture of Gaussian and/or non-Gaussian distributions  \citep{Hanson2002, Caudill2012, Karlsson2014, Garay2017, zeller2019finite}. \cite{Wang2019} proposed a mixture of factor analyzers for multivariate data that simultaneously performs clustering and dimension reduction, and \cite{Wang2021} extended it to predictors and censoring. 

Our primary contribution is an analysis of the Emory ADRC/HBS dataset using a censored multivariate Gaussian mixture of regressions. We implement an EM algorithm to address the important gap that software does not exist for the censored multivariate Gaussian mixture of regressions. We also address gaps in the current literature regarding the use of Wald tests of significance of the predictors in this context, wherein we approximate the information matrix using the empirical complete data score function. We also conduct a simulation study to address a gap regarding the impact of predictors and censoring on the accuracy of clustering. 

The remainder of this paper is organized as follows. In Section 2, we review the multivariate tobit model and describe the extension to censored Gaussian mixtures. We then describe an EM algorithm and method for inference. In  Section 3, we conduct simulations to illustrate the advantages of the censored multivariate GMR over methods ignoring or deleting the censored observations, and we also examine the selection of the number of clusters. In Section 4, we conduct an analysis of the Emory ADRC/HBS Dataset. Finally, we discuss findings and future research in Section 5.

\section{Modeling approach and estimation}
\label{sec:model}
In this section, we first review the multivariate tobit model and its estimation using an expectation-maximization (EM) algorithm. We then build upon this framework to derive an EM algorithm for the censored multivariate GMR.

\subsection{Multivariate censored regression (tobit model)}
\label{sec:tobit}
Let $\by_i$ be a $p$-dimensional random vector for the $i$th subject, $i=1,\dots,N$, which can be partitioned into two parts, 
\begin{linenomath}
    \begin{align*}\by_i=\begin{pmatrix}
                \by_{io_{i}}\\
                \by_{ic_{i}}
            \end{pmatrix},\; i\in \{1,2,...,N\},
    \end{align*}
\end{linenomath}
where $\by_{io_{i}}$ and $\by_{ic_{i}}$ denote the uncensored and censored dimensions of $\by_i$. We use a vector of censoring indicators $\bc_i$ to represent whether or not one particular dimension of $\by_i$ is censored, and its censoring directions are observed through $\bc_i=(c_{i1},...,c_{ij},...,c_{ip})^\top$, where $^\top$ denotes the transpose, such that:
\begin{linenomath}
\begin{align*}
c_{ij}=  
     \begin{cases}
       1, & \mathrm{Right-censored},\\
       0, & \mathrm{Uncensored}, \\
       -1,& \mathrm{Left-censored}.\\
     \end{cases}
\end{align*}
\end{linenomath}
Though the true values are not observed before they are censored, we can nevertheless further assume $\by_i$ are generated from the partially unobserved truth, as a latent random vector $\by_i^*=(y_{i1}^*,...,y_{ij}^*,...,y_{ip}^*)^\top$ with some known lower and upper detection limits:
\begin{linenomath}
\begin{align*}
    y_{ij}=\begin{cases}
         L_{j}\qquad    \mathrm{if}\qquad y_{ij}^*\le L_{j},  \\
         y_{ij}^*\qquad \mathrm{if}\qquad L_{j}<y_{ij}^*<U_{j},\\
         U_{j}\qquad    \mathrm{if}\qquad y_{ij}^*\ge U_{j}.
        \end{cases}
\end{align*}
\end{linenomath}
Similarly, we can partition $\by_i^*$ into observed and censored parts: 
\begin{linenomath}
\begin{align*}\by_i^*=\begin{pmatrix}
                \by_{io_{i}}\\
                \by_{ic_{i}}^*
            \end{pmatrix},\; i\in \{1,2,...,N\},
\end{align*}
\end{linenomath}
where $\by_{ic_{i}}^*$ are unobserved and censored as $\by_{ic_{i}}$. In a multivariate regression model setting, we also assume $\bx_i$ is a $d$-dimensional vector that represents the observed predictors of the $i$th subject. Then, the model can be formed as 
\begin{linenomath}
\begin{align}
    \by_i^*=\bbeta^\top\bx_i+\bepsilon_i,\label{eq:model_tobit}
\end{align}
\end{linenomath}
where $\bbeta$ is a $d\times p$ coefficient matrix and a primary parameter of interest, $\bepsilon_i$ is a $p$-dimensional vector of random noise and  $\bepsilon_i\overset{i.i.d.}{\sim} \mathcal{N}(\mathbf{0},\bSigma)$. 

Let $\bpsi$ denote the collection of parameters $\bbeta$ and $\bSigma$. Let $\bY$ be the $N \times p$ matrix of stacked observations $\by_i^\top$, $\bC$ be the $N \times p$ matrix of stacked censoring directions $\bc_i^\top$; and $\bX$ the $N \times d$ matrix of stacked predictor vectors $\bx_i^\top$. Then $\bpsi$ can be estimated from maximization of the incomplete data likelihood function:
\begin{linenomath}
\begin{align}
    \mathcal{L}(\bY;\bC,\bX,\bpsi)=\prod_{i=1}^Nf_{\by_{io_{i}}}(\by_{io_{i}};\bx_i, \bpsi)\int_{\mathcal{D}(\bc_i)} f_{\by_{ic_{i}}^*|\by_{io_{i}}}(\by_{ic_{i}};\bx_i,\bpsi),\label{eq:incomplete_lik_tobit}
\end{align}
\end{linenomath}
where $\mathcal{D}(\bc_i)$ is a domain in $\mathbb{R}^p$, depending on the censored patterns represented by $\bc_{i}$, and $f_{\by_{ic_i}^*|\by_{io_i}}$ is the conditional Gaussian density of the unobserved responses $\by_{ic_i}^*$ that experience censoring given the observed responses without censoring $\by_{io_i}$. Typically, \eqref{eq:incomplete_lik_tobit} is maximized numerically by the Newton–Raphson method \citep{Amemiya1973}. Here, we outline the EM algorithm similar to \citep{Fair1977,Ruud1991}, which we will extend to Gaussian mixtures in Section \ref{sec:ourmodel}. Let $\bY^*$ denote the $N \times p$ matrix of true observations formed by stacking $\by_i^{*\top}$. Define the complete data likelihood assuming $\by_i^*$, $i=1,\dots,N$, are observed:
\begin{linenomath}
\begin{align*}
    \mathcal{L}_c(\bY^*;\bC,\bX,\bpsi)=\prod_{i=1}^N\frac{1}{\sqrt{2^p\pi^p|\bSigma|}}e^{-\frac{1}{2}(\by_i^*-\bbeta^\top\bx_i)^\top\bSigma^{-1}(\by_i^*-\bbeta^\top\bx_i)}.
\end{align*}
\end{linenomath}
The EM algorithm steps are derived by maximization of the conditional expectation of this complete data log-likelihood function, with details in the Appendix \ref{Supplement:1.1}.

\subsection{Censored multivariate Gaussian mixture of regressions}
\label{sec:ourmodel}
Let $G$ denote the number of clusters. Later, we examine the selection of $G$ using information criteria. Then we can expand the model in \eqref{eq:model_tobit} to a mixture model:
\begin{linenomath}
\begin{align}
    \by_i^*|g=\bbeta_g^\top\bx_i+\bepsilon_{ig},\ g\in \{1,..,G\}.\label{eq:model_ours}
\end{align}
\end{linenomath}
Therefore, $\by_i^*|\{g,\bx_i\}\sim \mathcal{N}(\bbeta_g^\top\bx_i,\bSigma_g)$. Assuming observations are i.i.d., define the incomplete data likelihood function:
\begin{linenomath}
\begin{align}
    \mathcal{L}(\bY;\bC,\bX,\bPsi)=\prod_{i=1}^N\sum_{g=1}^G\omega_gf_{\by_{io_{i}}}(\by_{io_{i}};\bx_i, \bpsi_g)\int_{\mathcal{D}(\bc_i)} f_{\by_{ic_{i}}^*|\by_{io_{i}}}(\by_{ic_{i}};\bx_i,\bpsi_g),\label{eq:incomplete_lik}
\end{align}
\end{linenomath}
where $\bpsi_g$ are the vectorized parameters for each component; $\omega_g\in (0,1)$ are the mixing proportions of each mixture component subject to constraint: $\sum_{g=1}^G \omega_g=1$; and $\bPsi$ is the overall parameter vector, such that $\bPsi=(\omega_1,...,\omega_{G-1},\bpsi_1,...,\bpsi_G )$. Unlike the previously described regression model, in a mixture model setting, the direct maximization of the above likelihood function is not possible. Thus, we utilize the EM algorithm for parameter estimation. Let $z_{ig}$ denote an indicator variable equal to one if the $i$th subject is in the $g$th group and zero otherwise, and let $\bZ$ denote the $N \times G$ matrix formed by stacking $[z_{i1},\dots,z_{iG}]^\top$. Assuming both $\bZ$ and $\bY^*$ are observed, we can write the complete data likelihood function:
\begin{linenomath}
\begin{align}
    \mathcal{L}_c(\bY^*, \bZ ;\bC,\bX,\bPsi)=\prod_{i=1}^N\prod_{g=1}^G \left\{ \frac{\omega_g}{\sqrt{2^p\pi^p|\bSigma_g|}}e^{-\frac{1}{2}(\by_i^*-\bbeta_g^\top\bx_i)^\top\bSigma_g^{-1}(\by_i^*-\bbeta_g^\top\bx_i)}\right\}^{z_{ig}}.\label{eq:complete_lik}
\end{align}
\end{linenomath}
The EM algorithm is initialized with values $\bPsi^{(0)}$ and then iterates between the E- and M-steps until convergence, as described here. Let $\bPsi^{(k)}$ denote the values of the parameters from the previous iteration. Let $Z_{ig}$ denote a random indicator variable equal to one if the $i$th subject is in the $g$th group and zero otherwise. Define $\langle z_{ig}\rangle \equiv \E_{\bPsi^{(k)}}(Z_{ig}|\by_i)$, where $\E_{\bPsi^{(k)}}$ denotes the expectation evaluated using $\bPsi^{(k)}$. Let $\langle{\by_i^*}\rangle_g\equiv \E_{\bPsi^{(k)}}(\by_i^*|\by_i,Z_{ig}=1)$ and $\langle {\by_i^*{\by_i^*}^\top}\rangle_g\equiv \E_{\bPsi^{(k)}}(\by_i^*{\by_i^*}^\top|\by_i,Z_{ig}=1)$. Taking the conditional expectation of the complete data log-likelihood function:
\begin{linenomath}
\begin{align}
\begin{split}
    \mathcal{Q}_c(\bPsi;\bPsi^{(k)})
    &\propto \sum_{i=1}^N\sum_{g=1}^G\langle z_{ig}\rangle\{\ln \omega_g-\frac{1}{2}\ln{|\bSigma_g|}-\frac{1}{2}\tr\left(\bSigma_g^{-1}\langle \by_i^*{\by_i^*}^\top\rangle_g\right)\\
    &\qquad\qquad+\tr\left(\bSigma_g^{-1}\bbeta_g^\top\bx_i\langle {\by_i^*}\rangle_g^\top\right)-\frac{1}{2}\tr\left(\bSigma_g^{-1}\bbeta_g^\top\bx_i\bx_i^\top\bbeta_g\right)\}
\end{split}
\end{align}
\end{linenomath}
Then the EM algorithm steps are:
\begin{itemize}
    \item \textbf{E-step}:
    \begin{linenomath}
    \begin{align}
    \langle z_{ig}\rangle &\equiv E_{\bPsi^{(k)}}(Z_{ig}|\by_i)=\frac{\omega_g^{(k)}f_g(\by_i;\bx_i,\bpsi_g^{(k)})}{\sum_{h=1}^{G}\omega_h^{(k)}f_h(\by_i;\bx_i,\bpsi_h^{(k)})}
    \end{align}
    \begin{align}
    \langle {\by_i^*}\rangle_g\equiv E_{\bPsi^{(k)}}(\by_i^*|\by_i,Z_{ig}=1)=
    \begin{pmatrix}
                \by_{io_{i}}\\
                \langle \by_{ic_{i}}^*|\by_{io_{i}} \rangle_g
            \end{pmatrix}
    \end{align}
    \begin{align}
    \langle {\by_i^*{\by_i^*}^\top}\rangle_g\equiv E_{\bPsi^{(k)}}(\by_i^*{\by_i^*}^\top|\by_i,Z_{ig}=1)=
    \begin{pmatrix}
                \by_{io_{i}}\by_{io_{i}}^\top & \by_{io_{i}}\langle\by_{ic_{i}}^*|\by_{io_{i}}\rangle_g^\top\\
                \langle\by_{ic_{i}}^*|\by_{io_{i}}\rangle_g \by_{io_{i}}^\top & \langle \by_{ic_{i}}^* {\by_{ic_{i}}^*}^\top|\by_{io_{i}} \rangle_g
            \end{pmatrix},
    \end{align}
    \end{linenomath}
\end{itemize}
where 
$\langle \by_{ic_{i}}^* {\by_{ic_{i}}^*}^\top|\by_{io_{i}}\rangle_g$ 
and 
$\langle \by_{ic_{i}}^*|\by_{io_{i}}\rangle_g$
are the first and second moments of truncated conditional Gaussian distribution. These moments are calculated using the R package MomTrunc \citep{Galarza2021}.
\begin{itemize}
    \item \textbf{M-step}:
    \begin{align}
 \widetilde{\omega}_g=\sum_{i=1}^N \langle z_{ig} \rangle/N
    \end{align}
    \begin{align}
    \begin{split}
    \widetilde{\bbeta}_g
    =(\bX^\top\boldsymbol{\mathcal{Z}}_g\bX)^{-1}\bX^\top\boldsymbol{\mathcal{Z}}_g\langle \bY^*\rangle_g
    \end{split}
    \end{align}
    \begin{align}
    \begin{split}
    \widetilde{\bSigma}_g&=\frac{\sum_{i=1}^N\langle z_{ig} \rangle[\langle\by_i^*{\by_i^*}^\top\rangle_g-\widetilde{\bbeta}_g^\top\bx_i\langle{\by_i^*}\rangle_g^\top-\langle\by_i^*\rangle_g\bx_i^\top\widetilde{\bbeta}_g+\widetilde{\bbeta}_g^\top\bx_i\bx_i^\top\widetilde{\bbeta}_g]}{\sum_{i=1}^N\langle z_{ig} \rangle}
    \end{split}
    \end{align}
\end{itemize}
where $\langle{\bY^*}\rangle_g$ is the $N \times p$ matrix of stacked conditional expectations $\langle{\by_i^*}\rangle_g$, and $\boldsymbol{\mathcal{Z}}_g=\diag(\langle z_{1g} \rangle,...,\langle z_{Ng} \rangle)$.  Additional details are in the Appendix \ref{Supplement:1.2}.

\subsection{Hypothesis testing of within-cluster effects}
\label{s:inf}
Unlike some MLE algorithms in which the information matrix is automatically extracted (e.g., Newton-Rhapson updates), the information matrix is not directly calculated in the EM algorithm. As an alternative, some authors use bootstrapping \citep{McLachlan2000,OHagan2019}, which is generally reliable but computationally expensive. Instead, we approximate the observed information matrix using the empirical complete data score function. 

Under mild regularity conditions and weak consistency of the MLE that is a global maximizer in the interior of the parameter space $\hatbPsi \in \mathrm{int}(\bTheta)$ such that  $\hatbPsi\overset{p}{\to}\bPsi_0\in \bTheta$, then: 
\begin{equation}
    \frac{\sum_{i=1}^N\bs_c(\hatbPsi;\bPsi^{(k)}=\hatbPsi,\by_i)\bs_c^\top(\hatbPsi;\bPsi^{(k)}=\hatbPsi,\by_i)}{N}\overset{p}{\to}\mathcal{I}(\bPsi_0)
\end{equation}\\
where $\bPsi_0$ is the true parameter vector; $\bTheta$ is the parameter space; $\bs_c(\bPsi;\bPsi^{(k)}, \by_i)\equiv\frac{\partial Q_c(\bPsi;\bPsi^{(k)}, \by_i)}{\partial \bPsi}=\frac{\partial E_{\bPsi^{(k)}}(\log \mathcal{L}_{c_i}|\by_i)}{\partial \bPsi}$ is the first-order derivative of the individual conditional expectation of the complete date log-likelihood with respect to the parameters of interest. For details, see equation 2.60 in \cite{McLachlan2000}. We then conduct Wald tests of the within-cluster effects. This approach avoids the computation of second-order partial derivatives and is computationally feasible. \cite{McLachlan2000} note that the sample size in mixture models has to be large for valid inference. Our data application has $N>3,\!000$. In \cref{sec:simulation}, we show the type-1 error rates are in general close to their nominal levels in a setting with $N=1,\!000$.

\section{Simulations}
\label{sec:simulation}
\subsection{Simulation design}
We examine the censored multivariate GMR and estimators using two simulation scenarios: mild censoring and severe censoring. In both scenarios, unobserved data with $N=1,\!000$ are first generated from a three-cluster mixture of regressions with bivariate responses $(\bY_1,\bY_2)$ following Gaussian distributions where the means are linear transformations of an intercept $\bbeta_{g,0}$ where $g\in \{1,2,3\}$ and three predictors $(\bX_1, \bX_2, \bX_3)$. The simulation design is detailed in Table \ref{Supplement:Table.1} and summarized here. The predictors are generated from a mean-zero multivariate Gaussian distribution with the covariance matrix equal to the sample correlation matrix based on three continuous demographic features from the Emory ADRC/HBS (age, education and Montreal cognitive assessment). The true parameters are described in \cref{Supplement:Table.1}. 

In Scenario I, a lower detection limit equal to 0 is applied to the first response variable, $\bY_1$,  leading to around 4.1\% of observations left-censored on $\bY_1$, while an upper detection limit of 30 is applied to $\bY_2$ leading to around 13.7\% of observations right-censored on $\bY_2$. This leads to censoring levels similar to those found in the real data set (\cref{sec:realdataanalysis}). In Scenario II, a lower detection limit of 2.5 is applied to $\bY_1$ and upper detection limit of 26.5 to $\bY_2$ leading to around 40.2\% of observations left-censored and 37.2\% of observations right-censored, respectively. For each scenario, we performed 101 simulations with results from the median error used in figures. Scatter plots of the unobserved truth and the censored data from an example simulation are visualized in Appendix Figure \ref{Supplement:Figure.1}. 

We compare the censored multivariate GMR to three approaches: a multivariate mixture of regressions ignoring censoring, i.e., treating censored observations in the same manner as uncensored and using all observations (ignore-censor GMR); a multivariate mixture of Gaussians in which censored observations are deleted (delete-censor GMR); and the censored multivariate Gaussian mixture model ignoring predictors (censored GMM) \citep{Lee2012}. We report the mean and standard deviation (SD) across simulations of the mixing proportions $\omega_1$, $\omega_2$ and $\omega_3$. We report the mean and SD of Frobenius errors for other parameters. We evaluate the overall clustering accuracy using the adjusted Rand index (ARI) comparing the unobserved true labels against our modeled labels, where each observation is assigned to the class that had highest $\E_{\widehat{\bPsi}} (Z_{ig}|\by_i)$. 

We also examine the type-1 error rates of the censored multivariate GMR from 500 simulations using the estimates corresponding to parameters in which the true coefficient values are equal to 0.

We also investigate the selection of the number of clusters. We fit the censored multivariate GMR for $g=1,...,6$. For each of the 101 replicates and each $g$, the model is randomly initialized 32 times, and the solution with highest likelihood among the set of converged solutions is selected. We then calculate the integrated completed likelihood criterion (ICL) \citep{Biernacki2000}.

\subsection{Simulation results}

In Scenario I, the mixing proportions from the censored multivariate GMR are unbiased with small standard deviation, the mixing proportions from the ignore-censor GMR are similar, while bias increases in the delete-censor GMR with an overestimation of the frequency of group 2 and underestimation in group 3, and the mixing proportions are highly inaccurate in the censored GMM (\cref{simu_res}). For the regression coefficients, the censored multivariate GMR shows greater accuracy than other approaches. The ignore-censor GMR and delete-censor GMR are particularly inaccurate for group 3, which has regression coefficients leading to greater censoring than groups 1 and 2 (\cref{simu_res}). For the covariance matrices, the censored multivariate GMR is considerably more accurate than other approaches.

\begin{figure}[H]
    \centering
    \includegraphics[width=\textwidth]{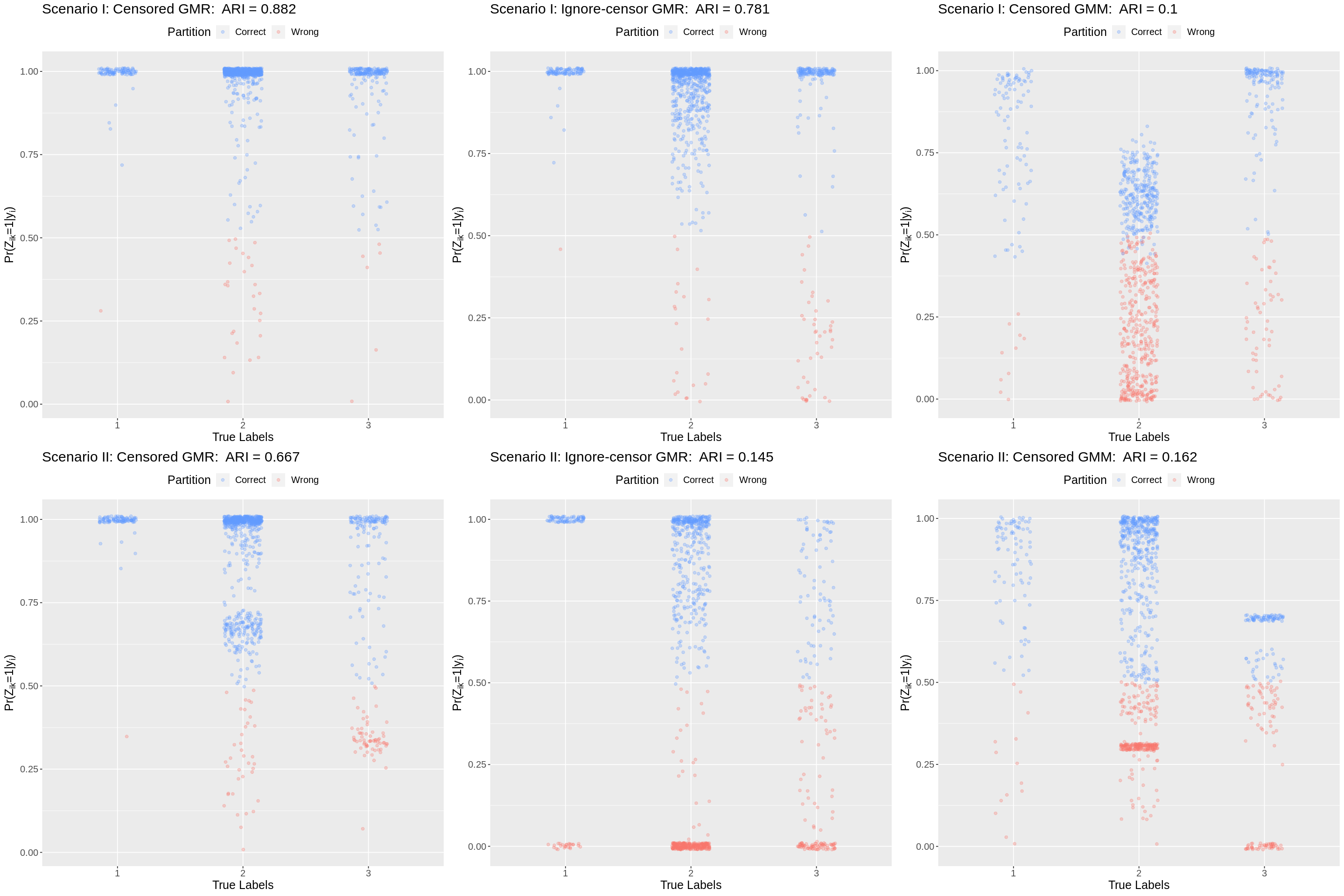}
    \caption{Clustering accuracy of the censored multivariate GMR, ignore-censoring multivariate GMR and censored mulivariate GMM. Displayed are the results associated with the median ARI from 101 simulations. Blue indicates that the observation was correctly classified using the posterior probabilities of cluster membership, and red indicates incorrect classification. Jitter added to improved visualization.}
    \label{fig:simfig}
\end{figure}

In Scenario II, we observe similar patterns but the benefits of the censored multivariate GMR are even greater (\cref{fig:simfig}, \cref{simu_res}). The censored multivariate GMR is still able to accurately estimate the mixing proportions, while ignore-censor GMR leads to gross inaccuracies. The censored GMM overestimated the frequency of group 2. Overall, the ARIs in Scenario II are lower than Scenario I, but the ARI from the censored multivariate GMR (0.68) is much higher than the ignore-censor GMR (0.08) and the censored GMM (0.17). Delete-censor GMR can not perform clustering on approximately 575 observations. The censored multivariate GMR still accurately estimates the regression coefficients and covariance matrices, while other approaches become highly inaccurate. 

These results highlight the need to model both the censoring and the predictors using the censored multivariate GMR. Even if a study is not interested in the influence of predictors, a mixture of regressions is necessary for accurate clustering. Moreover, the censoring must be modeled, otherwise both clustering and regression coefficient estimates are highly inaccurate. 

In Scenarios I and II, all type-1 error rates are near nominal levels, with the highest type-1 error rate equal to 0.056 (Appendix Table \ref{Supplement:Table.2}). 

In selecting the number of components in the censored multivariate GMR, ICL selects the correct number of components in both Scenarios I and II (Appendix Figure \ref{Supplement:Figure.2}).

\section{Real data application: Emory ADRC/HBS Dataset}
\label{sec:realdataanalysis}

\begin{figure}[H]
    \centering
    \includegraphics[width=\textwidth]{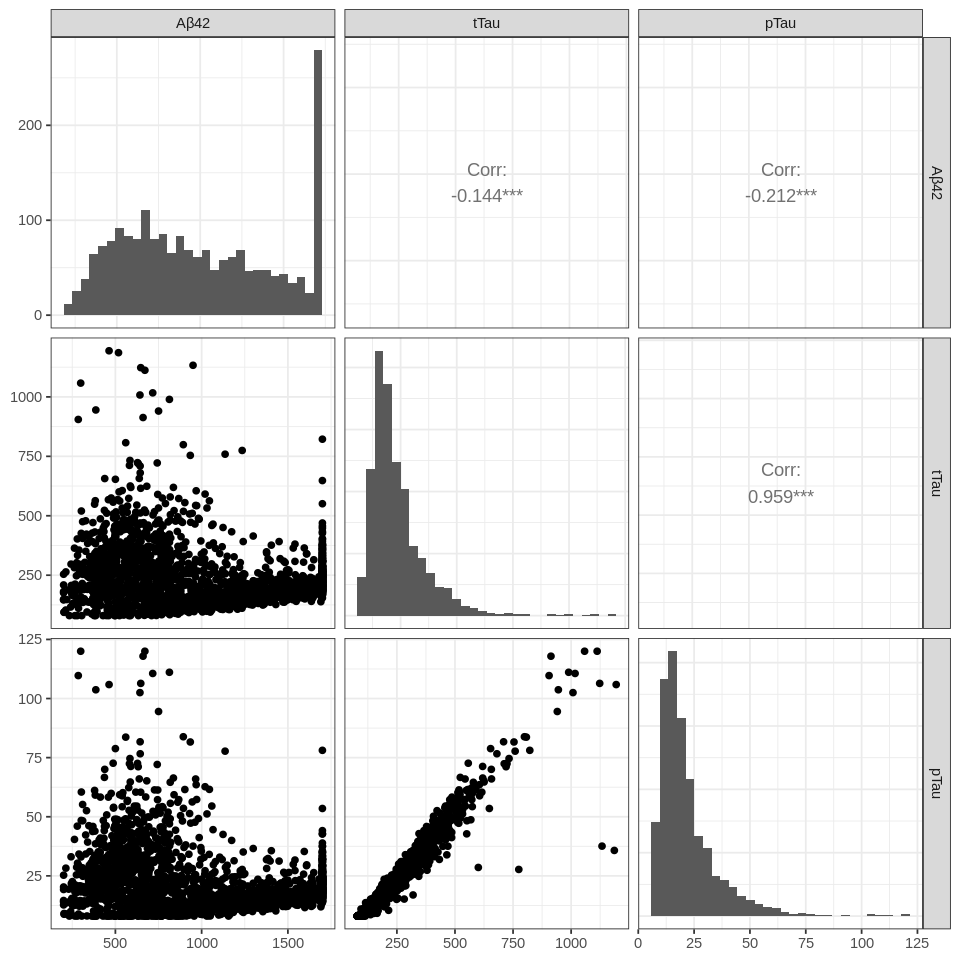}
    \caption{CSF biomarker data from the Emory ADRC/HBS Study.}
    \label{fig:00}
\end{figure}

The purposes of this analysis are twofold: 1) to identify patient clusters based upon their CSF biomarkers without utilizing possibly incorrect clinical diagnoses; and 2) to evaluate the within-cluster effects of the predictors on the CSF biomarkers. The Emory ADRC/HBS Dataset contains 3,004 individuals including 888 individuals with AD, 661 individuals with other cognitive disorders (Other) and 1,455 individuals who are cognitively normal (CN). 
Diagnosis is based on a combination of clinical history, neuropsychological testing, blood tests, structural neuroimaging and CSF biomarkers. All individuals included in our analyses provided informed consent to participate in research protocols approved by the Emory University Institutional Review Board (EHBS IRB00080300,
ADRC IRB00079069, NeuCog IRB00078273, Vascular IRB00084414, CRIN IRB00024959). All CSF samples are assayed on the Roche Elecsys platform to measure levels of amyloid-beta peptide 1-42 (Abeta42), total tau protein (tTau) and phosphorylated tau protein (pTau). Because of the detection limits of the biometric assay, the CSF biomarkers are subject to censoring: 10/3,004 observations of Abeta42 are left censored and 349/3,004 observations are right censored; 31/3,004 observations of tTau are left censored and 4 are right censored; and 110/3,004 observations of pTau are left-censored and 5 are right censored (\cref{fig:00}, \cref{real_demo}). Lower CSF Abeta42 corresponds to higher brain Abeta42 burden, such that we expect Abeta42 to be lower in patients with AD. In contrast, higher CSF tTau and pTau are associated with higher tau in the brain, and thus we expect tTau and pTau to be higher in patients with AD. A common approach is to use the ratio of Abeta42/tTau or Abeta42/pTau, where small values indicate AD pathology \citep{Hampel2008, Meyer2010}. However, using ratios obscures the censoring and discards information available from the multivariate approach. Since ignoring censoring led to erroneous clustering in our simulations, we believe the multivariate approach with the censored multivariate GMR will better reflect the underlying biology.

The data include age (decades), education level (decades), self-reported race, sex and apolipoprotein gene type. Due to small samples sizes in American Indian or Alaska Native ($n=6$), Asian ($n=36$), and Native Hawaiian or Other Pacific Islander ($n=7$), we created a binary variable for race equal to one if the participant was African American and zero otherwise. Additionally, there were 82 self-report Hispanic or Latino participants, which was not examined due to sample size. We included four levels: negative for the $\varepsilon4$ allele, heterozygous ApoE4 (ApoE4-1), homozygous ApoE4 (ApoE4-2) and missing data (\cref{real_demo}).  All individuals provided informed consent and all procedures are approved by the Emory University Institutional Review Board.

To select the optimal number of clusters, we calculate the ICL for the number of clusters $G=1,...,6$. For each $G$, we estimate an intercept and regression coefficients for the five demographic variables for the three CSF biomarkers, and we estimate the $3\times3$ covariance matrices. The model is randomly initialized 50 times and the solution with highest likelihood among the set of converge solutions is selected. (50/50 iterations converge for G=1 to 4, 45/50 for G=5, and 35/50 for G=6.)

The optimal number of clusters identified by ICL is equal to three (Appendix Figure \ref{Supplement:Figure.3}). 

\begin{figure}[H]
    \centering
    \includegraphics[width=\textwidth]{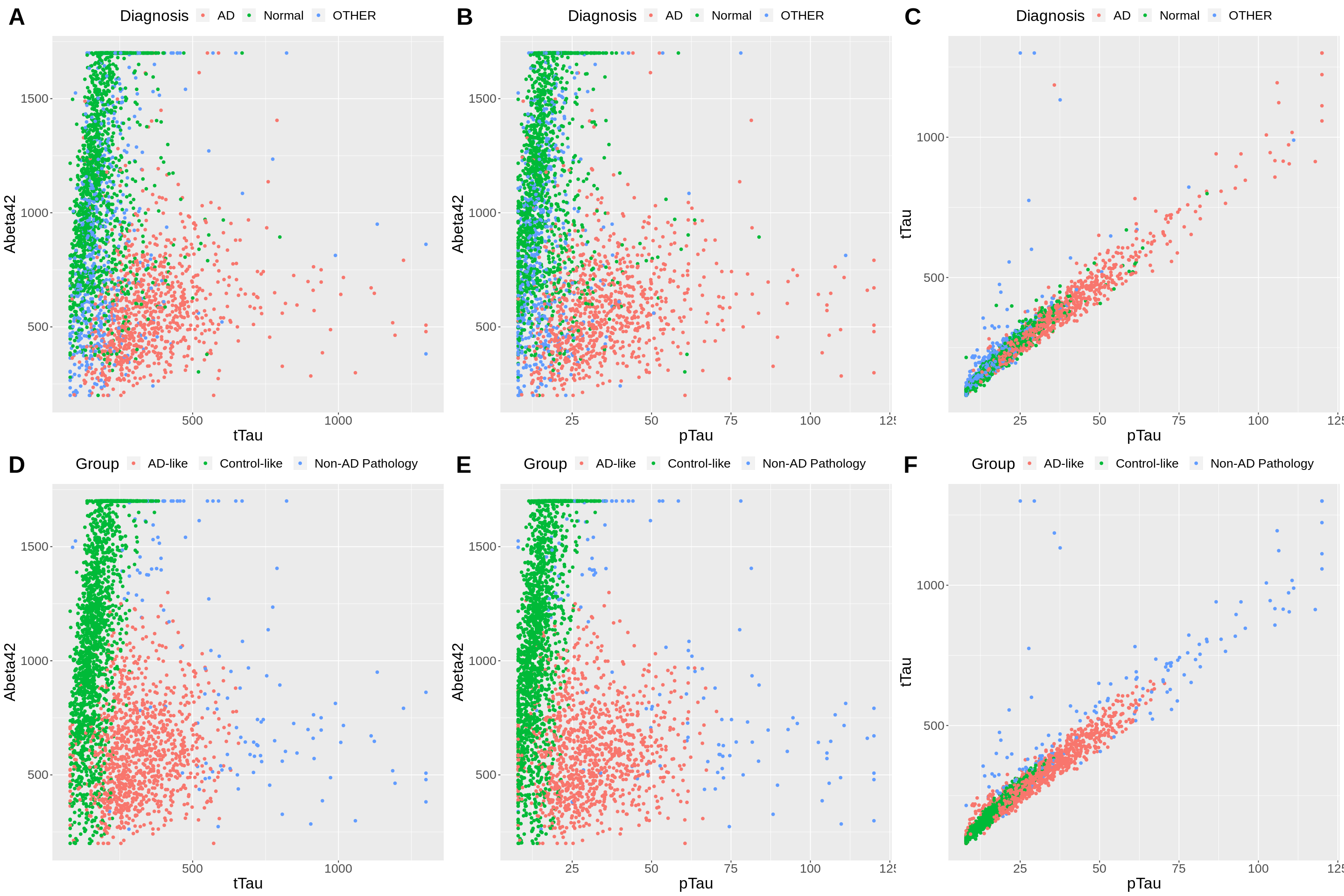}
    \caption{Scatter plots of the diagnosis versus the clusters identified with $G=3$. The top row has points colored by the diagnosis labels. The bottom row utilizes the output of the censored multivariate GMR. The cluster descriptions ``AD-like,'' ``Control-like'' and ``Non-AD pathology'' were determined by inspecting the characteristics of each cluster; see text.}\label{real_scatter} 
\end{figure}

To gain insight into the meaning of these groups, we visualize their relationship with the three (possibly incorrect) clinical diagnoses (\cref{real_scatter}). Panel A shows a scatterplot of Abeta42 and tTau colored by diagnosis group. Panel D shows the same scatterplot colored by the censored multivariate GMR clusters. Panels B and E show Abeta42 versus pTau for the AD-diagnosis and censored multivariate GMR, respectively, and Panels C and F show tTau versus pTau for AD-diagnosis and the censored multivariate GMR, respectively. 

We see that the AD labels in Panels A and B (pink) coincide with the pink censored multivariate GMR cluster in Panels D and E. Thus, we call the pink cluster in Panels D-F the ``AD-like pathology'' group.

The green group in Panels D-F \cref{real_scatter} tends to coincide with the ``Normal'' group in Panels A-C. 
We call this the ``control-like'' group. 

We call the blue group in Panels D-F of \cref{real_scatter} the ``Non-AD pathology'' group. The intercept in  \cref{real_est} indicates high CSF Abeta42 compared to control-like, i.e., low Abeta brain burden, which is generally considered non-AD pathology. However, the CSF tTau and pTau levels are higher than both the AD-like and control-like groups, indicating high tau brain burden, which may be associated with other types of dementia or neurological impairment. 

\begin{figure}[H]
    \centering
    \includegraphics[width=\textwidth]{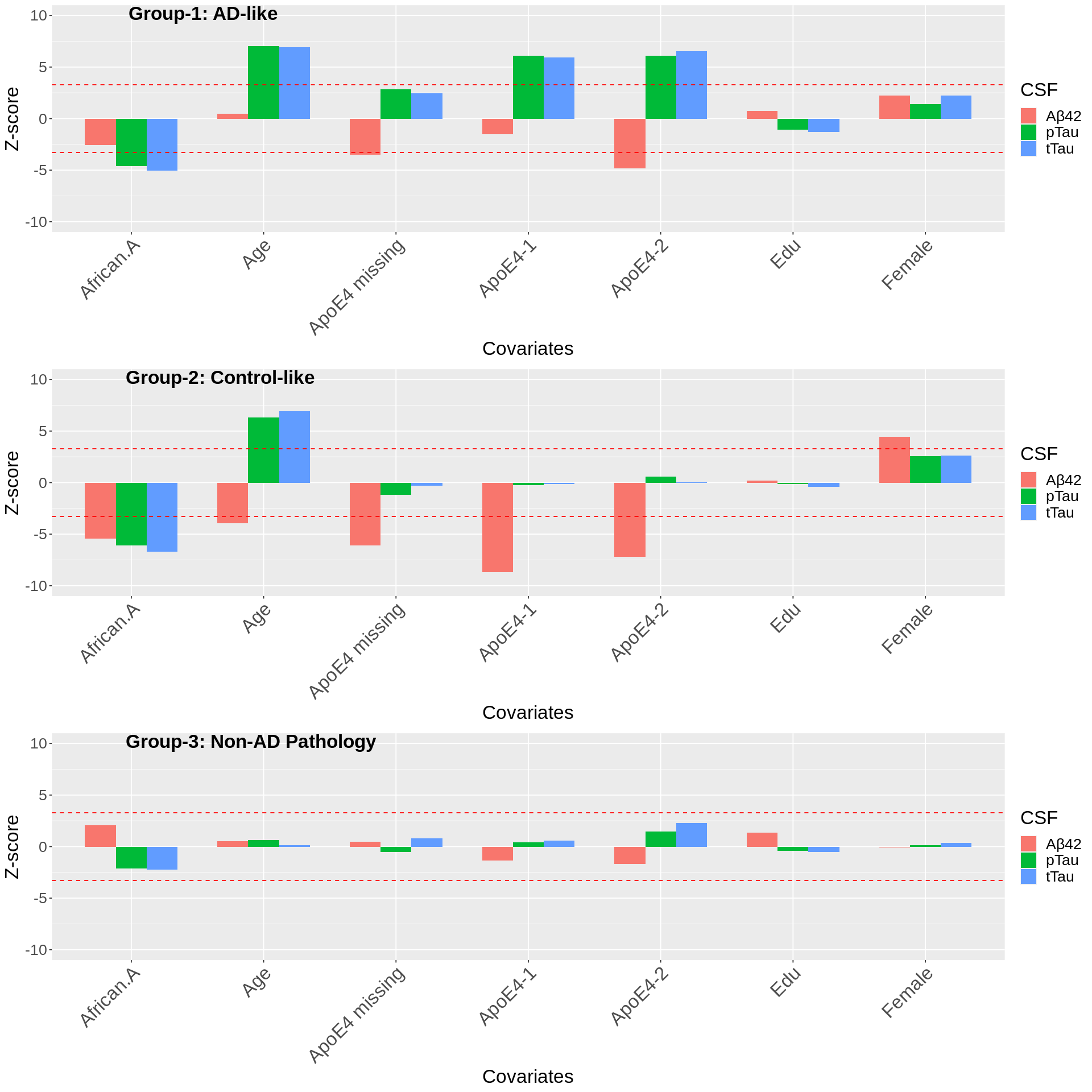}
    \caption{Bar plots of the Z-scores of the estimated within-cluster effects of the predictors in the selected $G=3$ model. The low Abeta42 high tau group has greatest overlap with AD diagnosis. The high Abeta42 low tau group has greatest overlap with cognitively normal. The high Abeta42 and high tau group has mixed pathology, which includes some MCI and other diagnosis. The dashed horizontal lines are the critical Z-score subject to Bonferroni correction for 63 comparisons.}\label{real_zscore}
\end{figure}

\underline{Group 1: AD-like pathology.}  
Abeta42 levels in African American participants are similar to the group composed of other races, while tTau and pTau are significantly lower in African American participants (\cref{real_est}, \cref{real_zscore}). A low ratio of Abeta42 to tTau is often used to classify individuals as AD. In this AD group, the Abeta42/tTau ratio for an African American participant would be larger than other races, implying that the conventional ratio would potentially misclassify African American patients. The effect of being an African American individual on tTau and pTau is similar to decreasing age by 18 years (\cref{real_est}). From a clinical perspective, African American participants in this group likely have AD, yet have lower tau burden and may be less likely to be diagnosed with AD, which suggests tau may not be a good biomarker for AD among African American individuals. This could have large implications on conventional approaches to classifying AD using CSF biomarker ratios, since conventional approaches are primarily based on studies in which the participants are primarily of European descent. 

There are large effects of ApoE4 for all three biomarkers, where ApoE4 decreases CSF Abeta42 and increases pTau and tTau. The coefficients for ApoE4-2 are all greater in magnitude than ApoE4-1 (\cref{real_est}). Carriers of two copies of ApoE4 have much higher levels of tTau and pTau in the AD-like group compared to carriers of two copies of ApoE4 in the control-like group. The coefficient for APOE4 missing (patients in which these data were not collected) is larger than APOE4-1. Upon further investigation, the missingness in APOE4 is not random: approximately 40\% are diagnosed with AD and 38\% with other pathology, suggesting that the frequency of APOE4-2 may be elevated in this group relative to the observed frequencies.

Other notable findings in this group are no association between age and Abeta42, but a positive relationship with tTau and pTau. Compared to other groups, the coefficients of age on tTau and pTau are large, which reflects a faster progressing tau pathology in the AD group. 

\underline{Group 2: Control-like}. 
In contrast to the AD-like group,  CSF Abeta42 are significantly lower in African American participants compared to the group composed of other races in the control-like group. Total tau and pTau are also significantly decreased in African American participants, but the coefficients are smaller than in the AD-like group. This again suggests that AD pathology differs in African American participants, and underscores importance of the mixture of regressions approach. Additionally, females in the control-like group had significantly higher levels of CSF Abeta42 than males. 

Carriers of ApoE4 have greatly reduced CSF Abeta42, but in contrast to the AD-like group, the tau levels are unchanged. 
Previous studies that have found that ApoE4 is associated with Abeta42 but not tau in cognitively normal aging \citep{Morris2010}. 

In contrast to the ``AD-like pathology'' group, CSF Abeta42 decreases with age (leading to an increase in brain Abeta42) in the control-like group. Since the levels of CSF Abeta42 levels are much lower at baseline in the AD-like group, CSF Abeta42 levels in the control-like group are still higher than the AD-like group at higher ages. Total tau and pTau increase with age, but at a slower rate compared to the AD-like group, which reflects age-progressing tau pathology in the control-like group.

\underline{Group 3:  Non-AD pathology}. Overall, we do not see significant relationships between the predictors and CSF biomarkers in this group (\cref{real_est}, \cref{real_zscore}). This may be due in part to the smaller mixing proportion implying small sample size and imprecise coefficient estimates. 

\section{Discussion}
\label{sec:discuss}
We used a censored multivariate Gaussian mixture of regressions with a feasible EM algorithm to examine predictor impacts on subgroups in CSF biomarkers of Alzheimer's Disease. The approach is similar to estimating multivariate regression effects in which all predictors interact with a group variable, but here we also learn the group labels. Our approach simultaneously identifies clusters while allowing the effects of predictors to differ for different clusters for a multivariate outcome with detection limits. In contrast to intensive bootstrap methods, we approximate the asymptotic covariance matrix of the within-cluster effects $\bbeta_g$ using the empirical complete score function. Our simulations show that this approach adequately controls type-1 error rates in large samples ($n \approx 1,\!000$). In simulations with moderate (comparable to our data application) and severe censoring, we show that ignoring censored records, deleting censored records or ignoring predictors creates substantial inaccuracies. Our approach results in large improvements in both the accuracy of clustering and regression estimates. Our simulations add to the latent class literature by demonstrating that modeling both the censoring and the predictors are important for accurate clustering. 

Our analysis of the Emory ADRC/HBS Dataset using the censored multivariate GMR reveals new insights. We identify three clusters that tend to align with an AD-like group, a control-like group and a third group with undefined non-AD pathology. Predictor effects vary across clusters. African American participants in the AD-like group had less severe tTau and pTau pathology. CSF biomarkers typically use the ratio of Abeta42 to tau, but previous studies may have based such determinations from studies of non-Hispanic Whites \citep{Meyer2010}. Recently, some researchers reported potential racial differences in CSF biomarkers \citep{Morris2019,garrett2019racial}, which aligns with our findings. Additionally, the effects of ApoE4 on CSF biomarker levels differed between the AD-like and control-like groups, females had higher CSF Abeta42 than males in the control-like group, there were no age impacts on Abeta42 in the AD-like group but significant effects in the control-like group, and age impacts on pTau and tTau were greatest in the AD-like group.

We found a higher proportion of patients in the AD-like group than were diagnosed with AD. This is expected since there is a significant number of cognitively normal individuals with asymptomatic AD  \citep{Jansen2022PrevalenceSpectrum,Jansen2015PrevalenceMeta-analysis}. Likewise, the vast majority of those clinically diagnosed as ``Other" would be expected to fall into a non-AD or control-like multivariate CSF distribution. A benefit of the censored multivariate GMR is that it generates probabilities of membership in each cluster. Future research can create an interactive tool to allow a  clinician to enter a patient's data and obtain probabilities for membership in the AD-like, control-like, and non-AD pathology groups, which can complement existing approaches to diagnosing AD.

There are a number of limitations of our approach. Model selection and interpretation can be challenging with increasing number of response variables, predictors and groups. Penalized approaches may be helpful in higher dimensions \citep{Khalili2013,Xie2010}. Another avenue for future research is to consider an alternative approach that models the probability of latent class membership as a function of covariates using multinomial regression \citep{Jacobs1991}. In our approach, the predictors indirectly impact the posterior probabilities. This results in more interpretable effects, which can deepen our understanding of the impact of demographics, behavior and genetics on biomarkers in complex neurological disorders.

\section{Software}
\ifblind
Code used in Section~\ref{sec:simulation} is available at \url{https://github.com/} BLINDED
\else
Code used in Section~\ref{sec:simulation} is available at \url{https://github.com/GanzhongTian/CensGMR}. 
\fi

\section*{Disclosure statement}
\ifblind
Blinded
\else
J. Lah is a consultant for Roche Diagnostics.
\fi
\section*{Funding}

\ifblind
Blinded
\else
G.T. was supported by R01 AG055634 and R01 AG070937. B.B.R. was supported by R21 AG066970. J.H. was supported by R01 AG055634 and P50 AG025688. J.L. was supported by R01 AG070937 and P50 AG025688. Funding for materials for Roche Elecsys assays were supported by a Roche IIS grant RD004723 to J.L. Additional funding for this work was provided by a generous gift from the Goizueta Foundation.
\fi

\pagebreak

\section*{Table 1}

\begin{table}[H]
    \centering
    \scalebox{0.8}{
    \small
    \begin{tabular}{c|c|c|c|c}
    \toprule
    \textbf{Packages} & \textbf{Censored} & \textbf{Multivariate} & \textbf{Mixture} & \textbf{Truncated} \\
    & \textbf{Outcomes} & \textbf{Outcomes} & \textbf{of Regressions} & \textbf{Multivariate Normal} \\
    \midrule
    mclust   &{\color{red}\xmark}    &{\color{green}\cmark}  &{\color{red}\xmark} &{\color{red}\xmark} \\
    mixtools  &{\color{red}\xmark}    &{\color{green}\cmark}  &{\color{green}\cmark}  &{\color{red}\xmark} \\
    FlexMix   &{\color{red}\xmark}    &{\color{red}\xmark}    &{\color{green}\cmark}  &{\color{red}\xmark} \\
    SMNCensReg &{\color{green}\cmark}  &{\color{red}\xmark}    &{\color{red}\xmark}  &{\color{red}\xmark} \\
    CensMFM &{\color{green}\cmark}  &{\color{green}\cmark}  &{\color{red}\xmark}  &{\color{red}\xmark} \\
    poLCA  &{\color{red}\xmark}    &{\color{red}\xmark}    &{\color{green}\cmark}  &{\color{red}\xmark} \\
    CensMixReg &{\color{green}\cmark}  &{\color{red}\xmark}    &{\color{green}\cmark} &{\color{red}\xmark} \\
    fmm (Stata) &{\color{green}\cmark} & {\color{red}\xmark} & {\color{green}\cmark}  &{\color{red}\xmark} \\
    proc fmm (SAS)  &{\color{green}\cmark} & {\color{red}\xmark} & {\color{green}\cmark}  &{\color{red}\xmark} \\
    Latent GOLD 6.0 & {\color{green}\cmark} & {\color{green}\cmark} & {\color{green}\cmark}  &{\color{red}\xmark} \\
    \bottomrule
    \end{tabular}}
    \caption{List of finite mixture modeling software.  mclust:   \cite{Scrucca2016}; 
 mixtools: \cite{benaglia2010mixtools};    FlexMix: \cite{grun2008flexmix}; SMNCensReg: \cite{garay2013smncensreg}; CensMFM: \cite{de2020censmfm};  poLCA:  \cite{linzer2011}; CensMixReg: \cite{sanchez2015package}; Latent GOLD 6.0:  \cite{vermunt2021lg}.}\label{soft_list}
\end{table}    

\pagebreak

\section*{Table 2}

\begin{table}[h!]
\centering
    \begin{tabular}{ |p{4cm}||p{5.5cm}|p{6cm}|}
    \hline
    \multicolumn{3}{|c|}{Simulation Cases, $N=1,000$, $\bY=(\bY_1,\bY_2)$} \\
    \hline
    Parameters& \multicolumn{2}{|c|}{True Values} \\
    \hline
    $\bpi=(\omega_1,\omega_2,\omega_3)$   &\multicolumn{2}{|c|}{$(0.1,0.7,0.2)$}\\
    $\bbeta=(\bbeta_1,\bbeta_2,\bbeta_3)$   &\multicolumn{2}{|c|}{$\begin{bmatrix}\begin{pmatrix}
        2 & 20\\
        0 & -2\\
        0 & 0 \\
        0 & 0
       \end{pmatrix},
       \begin{pmatrix}
        3 & 25\\
        1 & -3\\
        0 & 0\\
        0 & 0
        \end{pmatrix},
        \begin{pmatrix}
        3.5 & 30\\
        2   & -5\\
        0 & 0\\
        0 & 0
        \end{pmatrix}\end{bmatrix}$}\\
        
    $\bSigma=(\bSigma_1,\bSigma_2,\bSigma_3)$   &\multicolumn{2}{|c|}{
            $\begin{bmatrix} \begin{pmatrix}
                    1   & 0.1\\
                    0.1 & 1
            \end{pmatrix},
            \begin{pmatrix}
            2 & 0.2\\
            0.2 & 0.5
            \end{pmatrix},
            \begin{pmatrix}
            0.5 & 0.3\\
            0.3   & 2            \end{pmatrix}\end{bmatrix}$}\\
            \\
    \hline
    & Scenario I & Scenario II\\
    \hline
    Detection Limits & $\bY_1\in(0,\infty)$, $\bY_2 \in (-\infty,30)$ & $\bY_1 \in (2.5,\infty)$, $\bY_2 \in (-\infty,26.5)$\\
    \hline
    $\bY_1$ Censored \% & left-censored $4.1\%$ & left-censored $40.2\%$\\
    \hline
    $\bY_2$ Censored \% & right-censored $13.7\%$ & right-censored $37.2\%$\\
    \hline
    \end{tabular}
    \caption{Summary of simulation scenarios.}\label{Supplement:Table.1}
\end{table}

\pagebreak

\section*{Table 3}
\begin{table}[H]
    \centering
    \scalebox{0.8}{
    \begin{tabular}{l l l l l l l}
    \toprule
    \midrule
    \textbf{Scenario I} & & & & & &\\
    \midrule
    \textbf{Parameters}     &   \textbf{Truth}  &   \textbf{Censored GMR}  &   \textbf{Ignore-censor GMR}  &   \textbf{Delete-censor GMR}  &    \textbf{Censored GMM}  &\\
    \midrule
    $N$ &   -    &   1000   &   1000   &     852   &  1000  &\\
    \midrule
    $\omega_1$ &   0.10    &   \textbf{0.10(0.00)}   &   0.10(0.00)   &     0.12(0.00)   &  0.16(0.11)  &\\
    $\omega_2$ &   0.70    &   \textbf{0.70(0.01)}   &   0.72(0.02)   &     0.77(0.01)   &  0.56(0.18)  &\\
    $\omega_3$ &   0.20    &   \textbf{0.20(0.01)}   &   0.18(0.02)   &     0.11(0.02)   &  0.29(0.17) &\\
    \midrule
    $||\bbeta_1-\hatbbeta_1||_F$ & - &  \textbf{0.44(0.14)}   &   0.53(0.19)   &  0.44(0.15)   &  -  &\\
    $||\bbeta_2-\hatbbeta_2||_F$ & - &  \textbf{0.19(0.09)}   &   0.26(0.07)   &  0.23(0.08)   &  -  &\\
    $||\bbeta_3-\hatbbeta_3||_F$ & - &  \textbf{0.53(0.26)}   &   3.32(0.37)   &  1.61(0.83)   &  -  &\\
    $||\bSigma_1-\hatbSigma_1||_F$  & - &  \textbf{0.24(0.09)}   &   0.42(0.24)   &  0.28(0.10)   &  7.24(10.32)  &\\
    $||\bSigma_2-\hatbSigma_2||_F$  & - &  \textbf{0.11(0.06)}   &   0.16(0.06)   &  0.21(0.08)   &  8.34(2.74)  &\\
    $||\bSigma_3-\hatbSigma_3||_F$  & - &  \textbf{0.35(0.22)}   &   0.71(0.28)   &  0.70(0.39)   &  18.20(13.16)  &\\
    $||\bPsi-\hatbPsi||_F$       & - &  \textbf{0.88(0.23)}   &   3.48(0.34)   &  1.91(0.81)   &  -  &\\
    \midrule
    ARI       & 1 &  \textbf{0.89(0.02)}   &   0.82(0.03)   &  -   &    0.31(0.15)   &\\
    \midrule
    \midrule
    \textbf{Scenario II} & & & & & &\\
    \midrule
    \textbf{Parameters}     &   \textbf{Truth}  &   \textbf{Censored GMR}  &   \textbf{Ignore-censor GMR}  &   \textbf{Delete-censor GMR}  &    \textbf{Censored GMM}  &\\
    \midrule
    $N$ &   -    &   \textbf{1000}   &   1000   &     418   &  1000  &\\
    \midrule
    $\omega_1$ &   0.10    &   \textbf{0.10(0.00) }  &   0.39(0.06)   &     0.08(0.02)   &  0.15(0.08)  &\\
    $\omega_2$ &   0.70    &   \textbf{0.70(0.02)}   &   0.42(0.04)   &     0.73(0.06)   &  0.53(0.18) &\\
    $\omega_3$ &   0.20    &   \textbf{0.20(0.02)}   &   0.19(0.07)   &     0.19(0.06)   &  0.32(0.18) &\\
    \midrule
    $||\bbeta_1-\hatbbeta_1||_F$ & - &  \textbf{0.52(0.18)}   &   3.51(0.32)   &  1.52(0.42)   & -   &\\
    $||\bbeta_2-\hatbbeta_2||_F$ & - &  \textbf{0.23(0.09)}   &   1.21(0.26)   &  0.92(0.14)   & -   &\\
    $||\bbeta_3-\hatbbeta_3||_F$ & - &  \textbf{0.84(0.43)}   &   6.03(0.71)   &  4.81(1.20)   & -   &\\
    $||\bSigma_1-\hatbSigma_1||_F$  & - &  \textbf{0.40(0.21)}   &   3.06(0.47)   &  1.11(0.65)   & 23.75(94.17)   &\\
    $||\bSigma_2-\hatbSigma_2||_F$  & - &  \textbf{0.14(0.08)}   &   1.80(0.53)   &  0.95(0.18)   & 6.53(2.95)   &\\
    $||\bSigma_3-\hatbSigma_3||_F$  & - &  \textbf{0.59(0.40)}   &   2.40(1.30)   &  0.79(0.40)   & 4.57(4.13)   &\\
    $||\bPsi-\hatbPsi||_F$       & - &  \textbf{1.32(0.43)}   &   8.42(0.74)   &  5.49(1.01)   & -   &\\
    \midrule
    ARI       & 1 &  \textbf{0.68(0.03)}   &   0.12(0.04)   &  -   &    0.17(0.07)   &\\
    \bottomrule
    \end{tabular}
    }
    \caption{ Simulation results in Scenarios I and II from the censored multivariate GMR and three other approaches.  Ignore-censor multivariate GMR uses the mixture of Gaussian regressions while treating the censored data in the same manner as uncensored. Delete-censor multivariate GMR uses the mixture of Gaussian regressions but deleting censored observations. Censored multivariate GMM uses the censored mixture of Gaussians without considering the effects of predictors, consequently only $\omega_g$, $\bSigma_g$ and the ARI are comparable to other approaches. Reported are the mean (sd) estimate across simulations for $\omega_g$, the mean (sd) of Frobenius errors and adjusted Rand Index (ARI) from 101 replicates. The clustering results for the replicate associated with the median error is in Figure \ref{fig:simfig}. ARI is not reported for delete-censor multivariate GMR because it can not perform clustering on the censored observations.}\label{simu_res}
\end{table}
\pagebreak

\section*{Table 4}
\begin{table}[H]
\centering
        \scalebox{1}{
        \begin{tabular}{l l l l}
        \toprule
        \textbf{Variable} &    & \textbf{N=3,004}\\
        \midrule
        Abeta42: &&\\
        &$\# <200$: & 10\\
        &uncensored: & 869.4 (587.2,1291.0)\\
        &$\# >1,\!700$: & 349\\
        tTau: &&\\
        &$\# <80$: & 31\\
        &uncensored: & 206.2 (153.7, 288.3)\\
        &$\# >1,\!300$: & 4\\
        pTau: &&\\
        &$\# <8$: & 110\\
        &uncensored: & 17.84 (13.10, 26.78)\\
        &$\# >120$: & 5\\
        \midrule
        Clinical Diagnosis: &&\\
        &AD: & 888 (0.30) \\
        &Other: & 661 (0.22)\\
        &CN: & 1,455 (0.48)\\
        \midrule
        Age (decades)     &   & $6.61\ (5.94,7.18)$\\
        \midrule
        Educ (decades)    &   & $1.60\ (1.40,1.80)$\\
        \midrule
        Race: &&\\
        &American Indian or Alaska Native: & 6\\
        &Asian: & 36\\
        &Black or African American: & 495\\
        &White: & 2,454\\
        &Native Hawaiian or Other Pacific Islander: & 7\\
        &Other: & 6\\
        \midrule
        Gender: &&\\
        &Female: & 1,788\\
        &Male: & 1,216\\
        \midrule
        ApoE4: &&\\
        &$\varepsilon4/\varepsilon4$: & 193 \\
        &$\varepsilon3/\varepsilon4$ or $\varepsilon2/\varepsilon4$: & 900 \\
        &$\varepsilon4$ Negative: & 1,520\\
        & missing data:& 391 \\
        \bottomrule
        \end{tabular}
        }
        \caption{Patient demographics of the Emory Goizueta Alzheimer's Disease Research Center and the Emory Healthy Brain Study Dataset.}\label{real_demo}
    \end{table}
    
\pagebreak 

\section*{Table 5}
\begin{table}[H]
\centering
\scalebox{1}{
\begin{tabular}{l l l l l}
\toprule
            &     Abeta42        &      tTau        &       pTau         &\\
\midrule
Group 1: AD-like   &                     &        $\omega_1=0.370$ &                       &\\  \midrule

Intercept	&	632.55*** 	&	260.22***	&	24.57***\\
Age	(decades) &	3.79 	&	29.65*** 	&	3.35***\\
Edu	(decades) &	18.84 	&	-18.77 &		-1.7\\
Female	&	31.6* 	&	18.26* 	&	1.29\\
African American	&	-59.1* 	&	-63.95***	&	-6.53***\\
Apoe4-2 &		-140.69***	&	87.35*** 	&	9.51***\\
Apoe4-1	&	-23.41 	&	53.68*** 	&	6.19***\\
Apoe4 missing	&	-72.22*** 	&	29.54* 	&	3.83** \\

\midrule
Group 2: Control-like &                     &        $\omega_2=0.577$ &                       &\\  
\midrule
Intercept	&	1284.74***	&	179.31***	&	15.43***\\
Age	(decades) &	-58.96*** 	&	11.76*** 	&	1***\\
Edu	(decades) &	9.84 	&	-2.28 	&	-0.07\\
Female	&	111.74*** 	&	7.45** 	&	0.66*\\
African American	&	-190.45***	&	-26.57***	&	-2.24***\\
Apoe4-2	&	-684.8*** 	&	0.32 	&	0.55\\
Apoe4-1	&	-260.1*** 	&	-0.56 	&	-0.07\\
Apoe4 missing	&	-222.31***	&	-1.33 	&	-0.45 \\
\midrule
Group 3: Non-AD pathology   &                       &        $\omega_3=0.053$  &                       &\\  
\midrule

Intercept	&1116.67***	&556.61*** 	&53.79***\\
Age	(decades) &28.33 	&4.54 	&1.98\\
Edu	(decades) &260.32 	&-47.64 	&-4.97\\
Female	&-9.66 	&22.83 	&0.78\\
African American	&350.04 	&-245.07 	&-23.48\\
Apoe4-2	&-627.41 	&281.46* 	&20.89\\
Apoe4-1	&-156.82 	&32.89 	&2.64\\
Apoe4 missing	&62.93 	&66.62 	&-5.24\\

\bottomrule
\end{tabular}}
\caption{Estimated coefficient matrices. With *: $p<0.05$; **: $p<0.01$; ***: $p<0.001$, uncorrected.}\label{real_est}
\end{table}
\pagebreak 

\section*{Figure 1}
\pagebreak

\section*{Figure 2}

\pagebreak

\section*{Figure 3}
\pagebreak

\section*{Figure 4}

\renewcommand{\thefigure}{S.\arabic{figure}}
\renewcommand{\thetable}{S.\arabic{table}}

\appendix

\section{Proposed Model}

\subsection{EM algorithm of the multivariate censored regression}
\label{Supplement:1.1}
Based upon the model definition in Section 2.1 of the main manuscript, the complete data likelihood assuming $\by_i^*$, $i=1,\dots,N$, are observed is
\begin{align*}
    \mathcal{L}_c(\bY^*;\textbf{C},\bX,\bpsi)=\prod_{i=1}^N\frac{1}{\sqrt{2^p\pi^p|\bSigma|}}e^{-\frac{1}{2}(\by_i^*-\bbeta^\top\bx_i)^\top\bSigma^{-1}(\by_i^*-\bbeta^\top\bx_i)}.
\end{align*}
Let $\langle \cdot \rangle$ denote an expectation conditioning on the observed data $\by\in \mathbb{R}^{n\times p}$, then the conditional expectation of the complete data log-likelihood function takes the form:
\begin{align*}
    \begin{split}
        \mathcal{Q}_c(\bpsi;\bpsi^{(k)},\by)
        &\propto\sum_{i=1}^N\{-\frac{1}{2}\ln{|\bSigma|}-\frac{1}{2}\tr\left(\bSigma^{-1}\langle\by_i^*{\by_i^*}^\top\rangle\right)+\tr\left(\bSigma^{-1}\bbeta^\top\bx_i\langle {\by_i^*}\rangle^\top\right)\\
        &\qquad-\frac{1}{2}\tr\left(\bSigma^{-1}\bbeta^\top\bx_i\bx_i^\top\bbeta\right)\}.
    \end{split}
\end{align*}
The EM algorithm steps are derived by maximization of the conditional expectation of $\mathcal{Q}_c(\bpsi;\bpsi^{(k)};\by)$:

\begin{itemize}
    \item  \textbf{E-step}, update the conditional expectations based on old parameter values:
\begin{align}
    \langle {\by_i^*} \rangle\equiv \E_{\bpsi^{(k)}}(\by_i^*|\by_i)=
    \begin{pmatrix}
                \by_{io_{i}}\\
                \langle \by_{ic_{i}}^*|\by_{io_{i}} \rangle
    \end{pmatrix}
\end{align}
\begin{align}
    \langle {\by_i^*{\by_i^*}^\top} \rangle\equiv \E_{\bpsi^{(k)}}(\by_i^*{\by_i^*}^\top|\by_i)=
    \begin{pmatrix}
                \by_{io_{i}}\by_{io_{i}}^\top & \by_{io_{i}}\langle\by_{ic_{i}}^*|\by_{io_{i}}\rangle^\top\\
                \langle\by_{ic_{i}}^*|\by_{io_{i}}\rangle \by_{io_{i}}^\top & \langle \by_{ic_{i}}^* {\by_{ic_{i}}^*}^\top|\by_{io_{i}} \rangle
    \end{pmatrix}
\end{align}
\end{itemize}
where
\begin{align*}
    \langle \by_{ic_{i}}^* {\by_{ic_{i}}^*}^\top|\by_{io_{i}}\rangle
    =\boldsymbol{\mu}_{c_i|o_i}^2({\bbeta^\top\bx_i},\bSigma; \mathcal{D}(\textbf{c}_i))
\end{align*}
\begin{align*}
    \langle \by_{ic_{i}}^*|\by_{io_{i}}\rangle
    =\boldsymbol{\mu}_{c_i|o_i}^1({\bbeta^\top\bx_i},\bSigma; \mathcal{D}(\textbf{c}_i))
\end{align*}
are the first and second moments of the multivariate truncated conditional Gaussian distribution subject to the truncation region $\mathcal{D}(\textbf{c}_i)$ which can be numerically computed using a quasi-Monte Carlo integration algorithm \citep{Genz2002,Genz2004}. 

The conditional complete data score function w.r.t the parameters of interest:
\begin{align*}
    \bS_c(\bbeta;\bpsi^{(k)},\by)&\equiv\frac{\partial\mathcal{Q}_c(\bpsi;\bpsi^{(k)},\by)}{\partial \boldsymbol{\bbeta}}\\
    &=\sum_{i=1}^N\bx_i\langle {\by_i^*}\rangle^\top\bSigma^{-1}-\bx_i\bx_i^\top\bbeta\bSigma^{-1}\\
    \bS_c(\bSigma;\bpsi^{(k)},\by)&\equiv\frac{\partial\mathcal{Q}_c(\bpsi;\bpsi^{(k)},\by)}{\partial \boldsymbol{\bSigma}}\\
    &=\sum_{i=1}^N-\frac{1}{2}\bSigma^{-1}+\frac{1}{2}\bSigma^{-1}\langle\by_i^*{\by_i^*}^\top\rangle\bSigma^{-1}-\bSigma^{-1}\langle{\by_i^*}\rangle\bx_i^\top\bbeta\bSigma^{-1}+\frac{1}{2}\bSigma^{-1}\bbeta^\top\bx_i\bx_i^\top\bbeta\bSigma^{-1}
\end{align*}
\begin{itemize}
    \item \textbf{M-step}, maximize $Q_c(\bpsi;\bpsi^{(k)},\by)$:
    \begin{itemize}
        \item Solve the conditional complete data score $\bS_c(\bbeta;\bpsi^{(k)},\by)$ at $\textbf{0}$:
        \begin{align}
        \begin{split}
            \widetilde{\bbeta} &=(\sum_{i=1}^N\bx_i\bx_i^\top)^{-1}(\sum_{i=1}^N\bx_i\langle\by_i^* \rangle^\top)\\
            &=(\bX^\top\bX)^{-1}\bX^\top\langle \bY^*\rangle
        \end{split}
        \end{align}
        \item Solve the conditional score $\bS_c(\bSigma;\bpsi^{(k)},\by)$ at $\textbf{0}$ and plug-in $\widetilde{\bbeta}$:
        \begin{align}
        \begin{split}
            \widetilde{\bSigma}           &=\frac{\sum_{i=1}^N\langle\by_i^*{\by_i^*}^\top\rangle-\widetilde{\bbeta}^\top\bx_i\langle{\by_i^*}\rangle^\top-\langle\by_i^*\rangle\bx_i^\top\widetilde{\bbeta}+\widetilde{\bbeta}^\top\bx_i\bx_i^\top\widetilde{\bbeta}}{N}\\
            &=\frac{\sum_{i=1}^N\langle(\by_i^*-\widetilde{\bbeta}^\top\bx_i)(\by_i^*-\widetilde{\bbeta}^\top\bx_i)^\top\rangle}{N}\\
            &=\frac{\langle (\bY^*-\bX\widetilde{\bbeta})^\top(\bY^*-\bX\widetilde{\bbeta}) \rangle}{N}
        \end{split}
        \end{align}
    \end{itemize}
\end{itemize}
where $\langle{\bY^*}\rangle_g$ is the $N \times p$ matrix of stacked conditional expectations $\langle{\by_i^*}\rangle_g$. The EM algorithm iterates between the \textbf{E} and \textbf{M} steps sequentially till convergence.

\subsection{EM algorithm of the censored multivariate GMR (tobit regression)}
\label{Supplement:1.2}
Similarly, based upon the model definition in Section 2.2, the complete data likelihood assuming both the latent component labels $\mathbf{z}_{ig}$ and latent data $\by_i^*$ are observed:
\begin{align}
    \mathcal{L}_c(\bY^*, \bZ ;\textbf{C},\bX,\bPsi)=\prod_{i=1}^N\prod_{g=1}^G \left\{ \frac{\omega_g}{\sqrt{2^p\pi^p|\bSigma_g|}}e^{-\frac{1}{2}(\by_i^*-\bbeta_g^\top\bx_i)^\top\bSigma_g^{-1}(\by_i^*-\bbeta_g^\top\bx_i)}\right\}^{z_{ig}}.
\end{align}
Again letting $\langle \cdot \rangle$ denote an expectation conditioning on the observed data $\by \in \mathbb{R}^{N\times p}$, then the conditional expectation of the complete data log-likelihood function takes the form:
\begin{align}
\begin{split}
    \mathcal{Q}_c(\bPsi;\bPsi^{(k)},\by)
    &\propto \sum_{i=1}^N\sum_{g=1}^G\langle z_{ig}\rangle\{\ln \omega_g-\frac{1}{2}\ln{|\bSigma_g|}-\frac{1}{2}\tr\left(\bSigma_g^{-1}\langle \by_i^*{\by_i^*}^\top\rangle_g\right)\\
    &\qquad\qquad+\tr\left(\bSigma_g^{-1}\bbeta_g^\top\bx_i\langle {\by_i^*}\rangle_g^\top\right)-\frac{1}{2}\tr\left(\bSigma_g^{-1}\bbeta_g^\top\bx_i\bx_i^\top\bbeta_g\right)\}.
\end{split}
\end{align}
The EM algorithm steps are derived by maximization of the conditional expectation of $\mathcal{Q}_c(\bPsi;\bPsi^{(k)},\by)$:

\begin{itemize}
    \item  \textbf{E-step}, update the conditional expectations based on old parameter values:
    
\begin{align}
    \langle z_{ig}\rangle &\equiv \E_{\bPsi^{(k)}}(Z_{ig}|\by_i)=P_{\bPsi^{(k)}}(Z_{ig}=1|\by_i)=\frac{\omega_g^{(k)}f_g(\by_i;\bx_i,\bpsi_g^{(k)})}{\sum_{h=1}^{G}\omega_h^{(k)}f_h(\by_i;\bx_i,\bpsi_h^{(k)})}
\end{align}
    
\begin{align}
    \langle {\by_i^*}\rangle_g\equiv \E_{\bpsi_g^{(k)}}(\by_i^*|\by_i)=
    \begin{pmatrix}
                \by_{io_{i}}\\
                \langle \by_{ic_{i}}^*|\by_{io_{i}} \rangle_g
            \end{pmatrix}
    \end{align}
    \begin{align}
    \langle {\by_i^*{\by_i^*}^\top}\rangle_g\equiv \E_{\bpsi_g^{(k)}}(\by_i^*{\by_i^*}^\top|\by_i)=
    \begin{pmatrix}
                \by_{io_{i}}\by_{io_{i}}^\top & \by_{io_{i}}\langle\by_{ic_{i}}^*|\by_{io_{i}}\rangle_g^\top\\
                \langle\by_{ic_{i}}^*|\by_{io_{i}}\rangle_g \by_{io_{i}}^\top & \langle \by_{ic_{i}}^* {\by_{ic_{i}}^*}^\top|\by_{io_{i}} \rangle_g
            \end{pmatrix}
\end{align}
\end{itemize}
where
\begin{align*}
    \langle \by_{ic_{i}}^* {\by_{ic_{i}}^*}^\top|\by_{io_{i}}\rangle_g
    =\boldsymbol{\mu}_{c_i|o_i}^2({\bbeta_g^\top\bx_i},\bSigma_g; \mathcal{D}(\textbf{c}_i))
\end{align*}
\begin{align*}
    \langle \by_{ic_{i}}^*|\by_{io_{i}}\rangle_g
    =\boldsymbol{\mu}_{c_i|o_i}^1({\bbeta_g^\top\bx_i},\bSigma_g; \mathcal{D}(\textbf{c}_i))
\end{align*}
are the first and second moments of truncated conditional Gaussian distribution of the $g$th mixture component subject to the truncation region $\mathcal{D}(\bc_i)$, which can be numerically computed using a quasi-Monte Carlo integration algorithm developed by \citep{Genz2002,Genz2004}. 

The conditional complete data score function with respect to the parameters of interest is

\begin{align*}
    \bS_c(\bbeta_g;\bPsi^{(k)},\by)&\equiv\frac{\partial\mathcal{Q}_c(\bPsi;\bPsi^{(k)},\by)}{\partial \bbeta_g}\\
    &=\sum_{i=1}^N\langle {\bz_{ig}}\rangle\bx_i\langle{\by_i^*}\rangle^\top\bSigma_g^{-1}-\langle {\bz_{ig}}\rangle\bx_i\bx_i^\top\bbeta_g\bSigma_g^{-1}\\
    \bS_c(\bSigma_g;\bPsi^{(k)},\by)&\equiv\frac{\partial\mathcal{Q}_c(\bPsi;\bPsi^{(k)},\by)}{\partial \bSigma_g}\\
    &=\sum_{i=1}^N\langle{\bz_{ig}}\rangle\left(-\frac{1}{2}\bSigma_g^{-1}+\frac{1}{2}\bSigma_g^{-1}\langle\by_i^*{\by_i^*}^\top\rangle\bSigma_g^{-1}-\bSigma_g^{-1}\langle{\by_i^*}\rangle\bx_i^\top\bbeta_g\bSigma_g^{-1}+\frac{1}{2}\bSigma_g^{-1}\bbeta_g^\top\bx_i\bx_i^\top\bbeta_g\bSigma_g^{-1}\right)
\end{align*}
\begin{itemize}
    \item \textbf{M-step}, maximize $Q_c(\bPsi;\bPsi^{(k)},\by)$:
    \begin{itemize}
        \item Update $\omega_g$ by solving Lagrange multiplier:
        \begin{align}
                \widetilde{\omega}_g=\sum_{i=1}^N \langle z_{ig} \rangle/N
        \end{align}
        
        \item Solve the conditional complete data score $\bS_c(\bbeta_g;\bPsi^{(k)},\by)$ at $\textbf{0}$:
        \begin{align}\label{eq:beta_hat}
        \begin{split}
                \widetilde{\bbeta}_g&=(\sum_{i=1}^N\langle z_{ig} \rangle\bx_i\bx_i^\top)^{-1}(\sum_{i=1}^N\langle z_{ig} \rangle\bx_i\langle\by_i^* \rangle_g^\top)\\
                &=(\bX^\top\boldsymbol{\mathcal{Z}}_g\bX)^{-1}\bX^\top\boldsymbol{\mathcal{Z}}_g\langle \bY^*\rangle_g
        \end{split}
        \end{align}
        \item Solve the conditional score $\bS_c(\bSigma_g;\bPsi^{(k)},\by)$ at $\textbf{0}$ and plug-in $\widetilde{\bbeta}_g$:
        \begin{align}
        \begin{split}
            \widetilde{\bSigma}_g&=\frac{\sum_{i=1}^N\langle z_{ig} \rangle\left(\langle\by_i^*{\by_i^*}^\top\rangle_g-\widetilde{\bbeta}_g^\top\textbf{x}_i\langle{\by_i^*}\rangle_g^\top-\langle\by_i^*\rangle_g\textbf{x}_i^\top\widetilde{\bbeta}_g+\widetilde{\bbeta}_g^\top\textbf{x}_i\textbf{x}_i^\top\widetilde{\bbeta}_g\right)}{\sum_{i=1}^N\langle z_{ig} \rangle}\\
            &=\langle(\bY^*-\textbf{X}\widetilde{\bbeta}_g)^\top\boldsymbol{\mathcal{Z}}_g(\bY^*-\textbf{X}\widetilde{\bbeta}_g) \rangle_g/(\widetilde{\omega}_gN)
        \end{split}
        \end{align}
    \end{itemize}
\end{itemize}
The EM algorithm iterates between the \textbf{E} and \textbf{M} steps sequentially until convergence.

\pagebreak

\section{Tables}

\begin{table}[h!]
\centering
\scalebox{1}{
    \begin{tabular}{ |p{4cm}||p{5cm}|p{6cm}|  }
    \hline
    \multicolumn{3}{|c|}{500 replications, $N=1,000$, $\bY=(\bY_1,\bY_2)$} \\
    \hline
    Parameters& \multicolumn{2}{|c|}{True Values} \\
    \hline
    $\bbeta=(\bbeta_1,\bbeta_2,\bbeta_3)$   &\multicolumn{2}{|c|}{$\begin{pmatrix}
        2 & 20\\
        0 & -2\\
        0 & 0 \\
        0 & 0
       \end{pmatrix},
       \begin{pmatrix}
        3 & 25\\
        1 & -3\\
        0 & 0 \\
        0 & 0
        \end{pmatrix},
        \begin{pmatrix}
        3.5 & 30\\
        2   & -5\\
        0 & 0 \\
        0 & 0
        \end{pmatrix}$}\\
    \hline
    Type-1 error rates of Scenario I:   &\multicolumn{2}{|c|}{$\begin{pmatrix}
    -	& -\\
    0.038	&-\\
    0.024	&0.044\\
    0.028	&0.040\\
       \end{pmatrix},
       \begin{pmatrix}
    -	&-\\
    -	&-\\
    0.040	&0.048\\
    0.042	&0.056\\
 
        \end{pmatrix},
        \begin{pmatrix}
    -	&-\\
    -&-\\
    0.048	&0.046\\
    0.046	&0.036\\
        \end{pmatrix}$}\\
    \hline
    Type-1 error rates of Scenario II:   &\multicolumn{2}{|c|}{$\begin{pmatrix}
    -   & -\\
    0.034   & -\\
    0.022   & 0.042\\
    0.024   & 0.050\\
       \end{pmatrix},
       \begin{pmatrix}
    -	& -\\
    -	& -\\
    0.038	& 0.042\\
    0.036	& 0.046\\
        \end{pmatrix},
        \begin{pmatrix}
    - &-\\
    -	&-\\
    0.038	&0.038\\
    0.048	&0.052\\
        \end{pmatrix}$}\\
    \hline
    \end{tabular}
}
    \caption{\label{Supplement:Table.2} Type-1 error rates for within-cluster effects in Scenario I (mild censoring) and Scenario II (severe censoring).}
\end{table}

\pagebreak    

\section{Figures}

\begin{figure}[H]
    \centering
    \includegraphics[width=\textwidth]{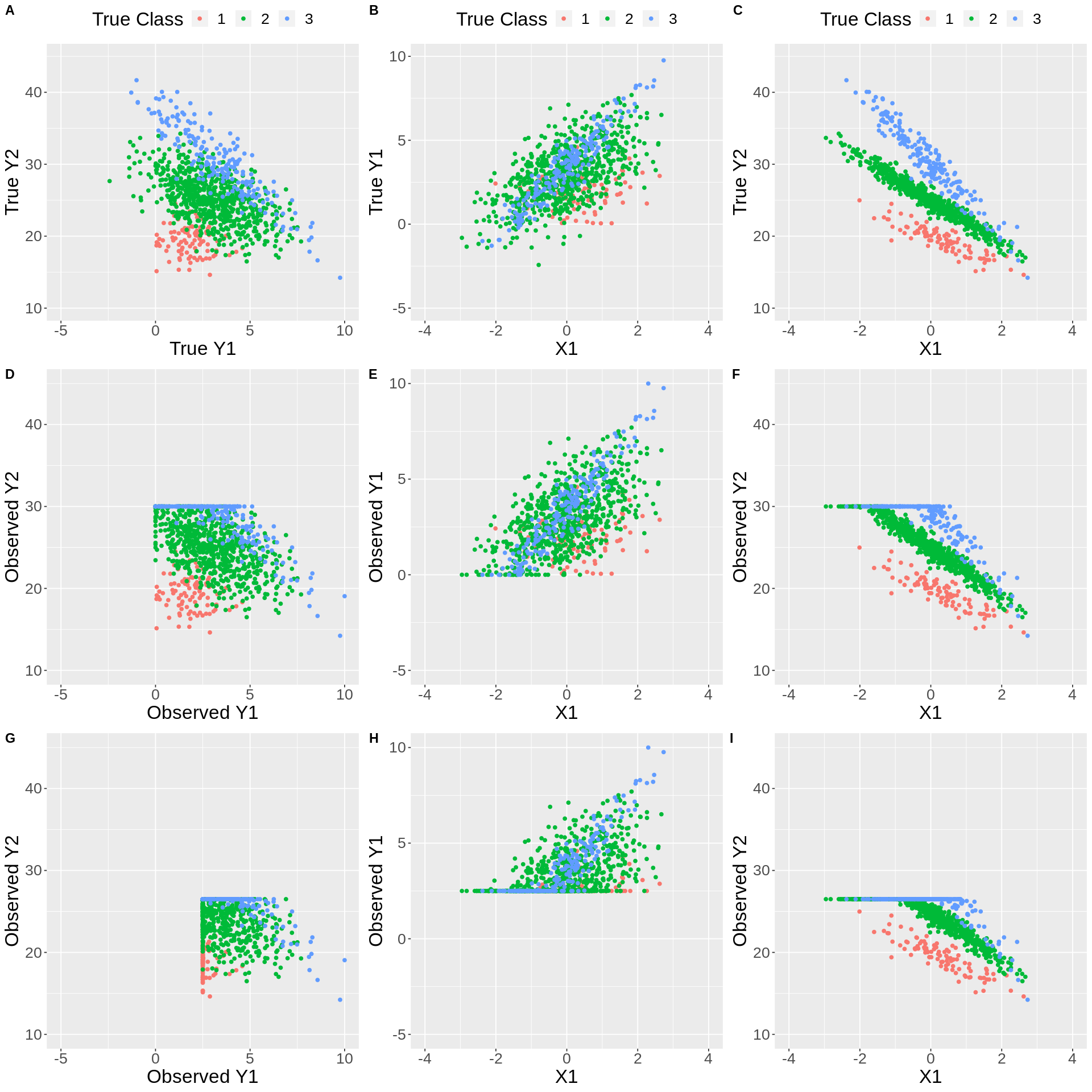}
    \caption{Example simulated data set and censoring mechanism. The data prior to censoring is depicted in row 1. In Scenario I (row 2), $\bY_1$ is left-censored at 0 and $\bY_2$ is right-censored at 30. In Scenario II (row 3), $\bY_1$ is left-censored at 2.5 and $\bY_2$ is right-censored at 30.}\label{Supplement:Figure.1}
    
\end{figure}

\begin{figure}[H]
    \centering
    \includegraphics[width=\textwidth]{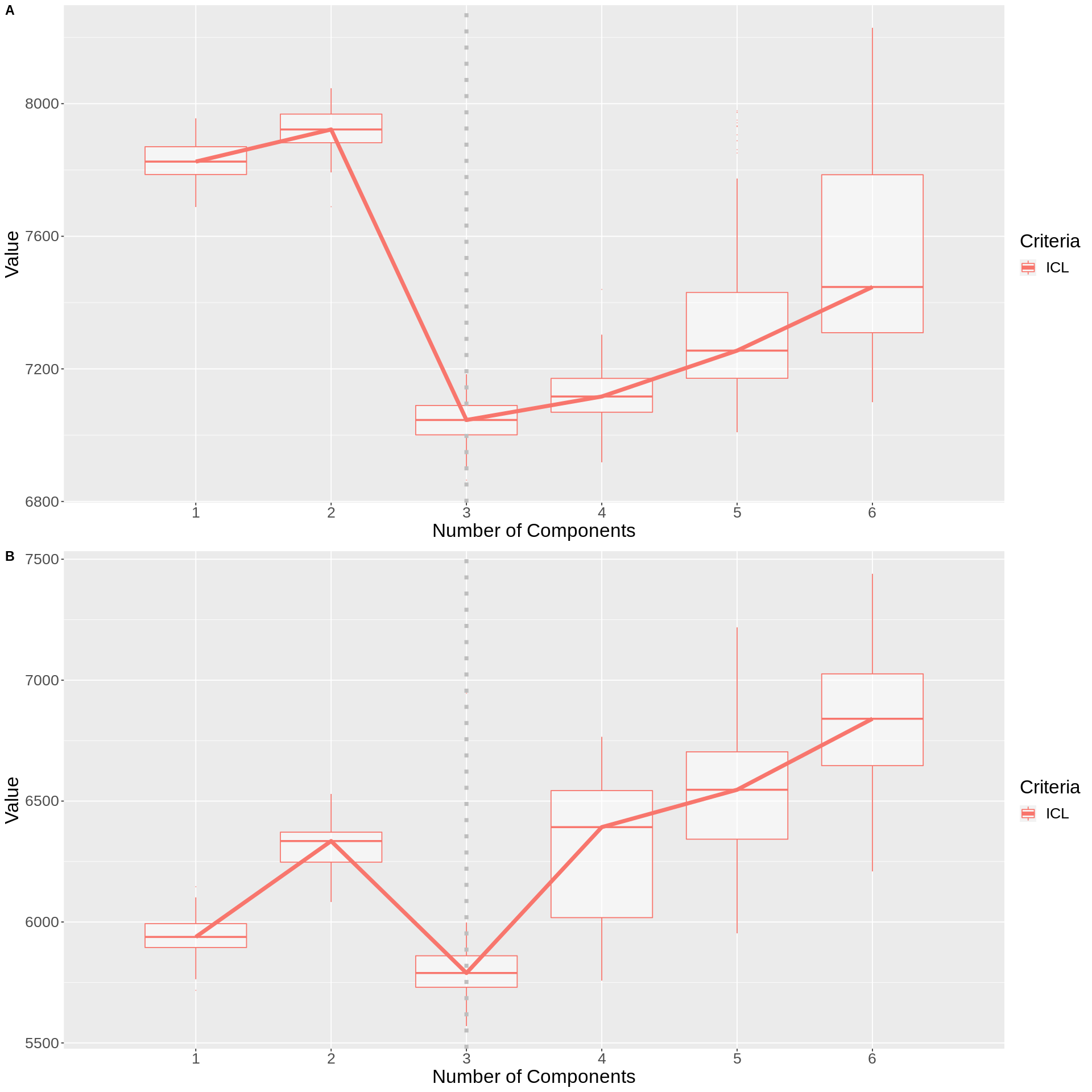}
    \caption{\label{Supplement:Figure.2} Selection of the optimal $G$ for Scenario I \& II on 101 data replicates. \textbf{A}: Selection of the optimal $G$ in Scenario I; \textbf{B}: Selection of the optimal $G$ in Scenario II.}
\end{figure}

\begin{figure}[H]
    \centering
    \includegraphics[width=\textwidth]{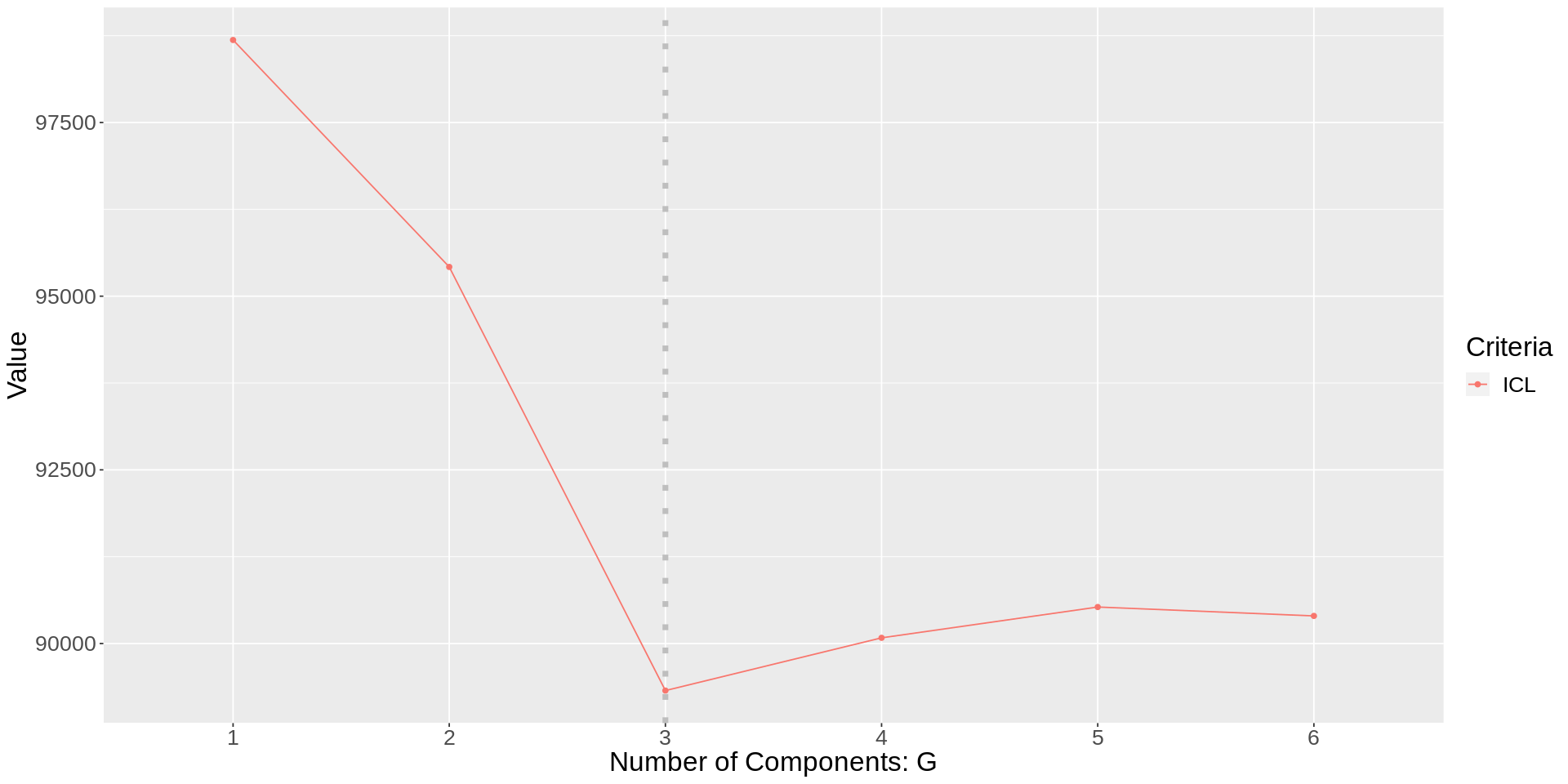}
    \caption{\label{Supplement:Figure.3} Selecting the optimal $G$ on the Emory Cognitive Neurology Biomarker Data. ICL indicates 3 groups.}
\end{figure}

\pagebreak

\bibliographystyle{apalike}
\bibliography{references}

\begin{thebibliography}{}

\bibitem[Amemiya, 1973]{Amemiya1973}
Amemiya, T. (1973).
\newblock {Regression Analysis when the Dependent Variable Is Truncated
  Normal}.
\newblock {\em Econometrica}, 41(6):997.

\bibitem[Benaglia et~al., 2010]{benaglia2010mixtools}
Benaglia, T., Chauveau, D., Hunter, D.~R., and Young, D.~S. (2010).
\newblock mixtools: an r package for analyzing mixture models.
\newblock {\em Journal of statistical software}, 32:1--29.

\bibitem[Biernacki et~al., 2000]{Biernacki2000}
Biernacki, C., Celeux, G., and Govaert, G. (2000).
\newblock {Assessing a mixture model for clustering with the integrated
  completed likelihood}.
\newblock {\em IEEE Transactions on Pattern Analysis and Machine Intelligence},
  22(7):719--725.

\bibitem[Blennow et~al., 2015]{Blennow2015}
Blennow, K., Dubois, B., Fagan, A.~M., Lewczuk, P., De~Leon, M.~J., and Hampel,
  H. (2015).
\newblock {Clinical utility of cerebrospinal fluid biomarkers in the diagnosis
  of early Alzheimer's disease}.
\newblock {\em Alzheimer's {\&} Dementia}, 11(1):58--69.

\bibitem[Caudill, 2012]{Caudill2012}
Caudill, S.~B. (2012).
\newblock {A partially adaptive estimator for the censored regression model
  based on a mixture of normal distributions}.
\newblock {\em Stat Methods Appl}, 21:121--137.

\bibitem[Coates and Ng, 2012]{Coates2012}
Coates, A. and Ng, A.~Y. (2012).
\newblock {Learning Feature Representations with K-Means}.
\newblock In {Montavon G.}, {Orr G.B.}, and {M{\"{u}}ller KR.}, editors, {\em
  Neural Networks: Tricks of the Trade}, chapter~22, pages 561--580. Springer,
  Berlin, Heidelberg, 2nd edition.

\bibitem[Collins and Huynh, 2014]{Collins2014}
Collins, J. and Huynh, M. (2014).
\newblock {Estimation of diagnostic test accuracy without full verification: a
  review of latent class methods}.
\newblock {\em Statistics in medicine}, 33(24):4141.

\bibitem[De~Alencar et~al., 2020]{de2020censmfm}
De~Alencar, F., Galarza, C., Matos, L., and Lachos, V. (2020).
\newblock Censmfm: Finite mixture of multivariate censored/missing data.
\newblock {\em R package version}, 2.

\bibitem[Dubois et~al., 2007]{Dubios2007b}
Dubois, B., Feldman, H.~H., Jacova, C., DeKosky, S.~T., Barberger-Gateau, P.,
  Cummings, J., et~al. (2007).
\newblock {Research criteria for the diagnosis of Alzheimer's disease: revising
  the NINCDS–ADRDA criteria}.
\newblock {\em The Lancet Neurology}, 6(8):734--746.

\bibitem[Fair, 1977]{Fair1977}
Fair, R.~C. (1977).
\newblock {A Note on the Computation of the Tobit Estimator}.
\newblock {\em Econometrica}, 45(7):1723.

\bibitem[Galarza et~al., 2021]{Galarza2021}
Galarza, C.~E., Kan, R., and Lachos, V.~H. (2021).
\newblock {MomTrunc: Moments of Folded and Doubly Truncated Multivariate
  Distributions}.
\newblock {\em R package version 5.97}.

\bibitem[Garay et~al., 2013]{garay2013smncensreg}
Garay, A., Lachos, V., and Massuia, M. (2013).
\newblock Smncensreg: Fitting univariate censored regression model under the
  scale mixture of normal distributions.
\newblock {\em R package version}, 2.

\bibitem[Garay et~al., 2017]{Garay2017}
Garay, A.~M., Lachos, V.~H., Bolfarine, H., and Cabral, C.~R. (2017).
\newblock {Linear censored regression models with scale mixtures of normal
  distributions}.
\newblock {\em Stat Papers}, 58:247--278.

\bibitem[Garrett et~al., 2019]{garrett2019racial}
Garrett, S.~L., McDaniel, D., Obideen, M., Trammell, A.~R., Shaw, L.~M.,
  Goldstein, F.~C., and Hajjar, I. (2019).
\newblock Racial disparity in cerebrospinal fluid amyloid and tau biomarkers
  and associated cutoffs for mild cognitive impairment.
\newblock {\em JAMA network open}, 2(12):e1917363--e1917363.

\bibitem[Genz, 2004]{Genz2004}
Genz, A. (2004).
\newblock {Numerical computation of rectangular bivariate and trivariate normal
  and t probabilities}.
\newblock {\em Statistics and Computing 2004 14:3}, 14(3):251--260.

\bibitem[Genz and Bretz, 2002]{Genz2002}
Genz, A. and Bretz, F. (2002).
\newblock {Comparison of methods for the computation of multivariate t
  probabilities}.
\newblock {\em Journal of Computational and Graphical Statistics},
  11(4):950--971.

\bibitem[Goetz et~al., 2019]{Goetz2019}
Goetz, M.~E., Hanfelt, J.~J., John, S.~E., Bergquist, S.~H., Loring, D.~W.,
  Quyyumi, A., Clifford, G.~D., Vaccarino, V., Goldstein, F., Johnson, T.~M.,
  Kuerston, R., Marcus, M., Levey, A.~I., and Lah, J.~J. (2019).
\newblock {Rationale and Design of the Emory Healthy Aging and Emory Healthy
  Brain Studies}.
\newblock {\em Neuroepidemiology}, 53(3-4):187--200.

\bibitem[Goldfeld and Quandt, 1973]{Goldfeld1973}
Goldfeld, S.~M. and Quandt, R.~E. (1973).
\newblock {A Markov model for switching regressions}.
\newblock {\em Journal of Econometrics}, 1(1):3--15.

\bibitem[Grun and Leisch, 2008]{grun2008flexmix}
Grun, B. and Leisch, F. (2008).
\newblock Flexmix version 2: finite mixtures with concomitant variables and
  varying and constant parameters.

\bibitem[Hampel et~al., 2008]{Hampel2008}
Hampel, H., B{\"{u}}rger, K., Teipel, S.~J., Bokde, A.~L., Zetterberg, H., and
  Blennow, K. (2008).
\newblock {Core candidate neurochemical and imaging biomarkers of Alzheimer’s
  disease}.
\newblock {\em Alzheimer's {\&} Dementia}, 4(1):38--48.

\bibitem[Hanson and Johnson, 2002]{Hanson2002}
Hanson, T. and Johnson, W.~O. (2002).
\newblock {Modeling regression error with a mixture of Polya trees}.
\newblock {\em Journal of the American Statistical Association},
  97(460):1020--1033.

\bibitem[Jacobs et~al., 1991]{Jacobs1991}
Jacobs, R.~A., Jordan, M.~I., Nowlan, S.~J., and Hinton, G.~E. (1991).
\newblock {Adaptive Mixtures of Local Experts}.
\newblock {\em Neural Computation}, 3(1):79--87.

\bibitem[Jansen et~al., 2022]{Jansen2022PrevalenceSpectrum}
Jansen, W.~J., Janssen, O., Tijms, B.~M., and Others (2022).
\newblock {Prevalence Estimates of Amyloid Abnormality Across the Alzheimer
  Disease Clinical Spectrum}.
\newblock {\em JAMA Neurology}, 79(3):228--243.

\bibitem[Jansen et~al., 2015]{Jansen2015PrevalenceMeta-analysis}
Jansen, W.~J., Ossenkoppele, R., Knol, D.~L., Tijms, B.~M., and Others (2015).
\newblock {Prevalence of Cerebral Amyloid Pathology in Persons Without
  Dementia: A Meta-analysis}.
\newblock {\em JAMA}, 313(19):1924--1938.

\bibitem[Jedidi et~al., 1993]{Jedidi1993}
Jedidi, K., Ramaswamy, V., and Desarbo, W.~S. (1993).
\newblock {A maximum likelihood method for latent class regression involving a
  censored dependent variable}.
\newblock {\em Psychometrika}, 58(3):375--394.

\bibitem[Karlsson and Laitila, 2014]{Karlsson2014}
Karlsson, M. and Laitila, T. (2014).
\newblock {Finite mixture modeling of censored regression models}.
\newblock {\em Stat Papers}, 55:627--642.

\bibitem[Khalili and Lin, 2013]{Khalili2013}
Khalili, A. and Lin, S. (2013).
\newblock {Regularization in Finite Mixture of Regression Models with Diverging
  Number of Parameters}.
\newblock {\em Biometrics}, 69(2):436--446.

\bibitem[Lee and Scott, 2012]{Lee2012}
Lee, G. and Scott, C. (2012).
\newblock {EM algorithms for multivariate Gaussian mixture models with
  truncated and censored data}.
\newblock {\em Computational Statistics {\&} Data Analysis}, 56(9):2816--2829.

\bibitem[Linzer and Lewis, 2011]{linzer2011}
Linzer, D.~A. and Lewis, J.~B. (2011).
\newblock {poLCA}: An {R} package for polytomous variable latent class
  analysis.
\newblock {\em Journal of Statistical Software}, 42(10):1--29.

\bibitem[McLachlan and Peel, 2000]{McLachlan2000}
McLachlan, G.~J. and Peel, D. (2000).
\newblock {\em {Finite mixture models}}.
\newblock Wiley.

\bibitem[Meyer et~al., 2010]{Meyer2010}
Meyer, G.~D., Shapiro, F., Vanderstichele, H., Vanmechelen, E., Engelborghs,
  S., Deyn, P. P.~D., Coart, E., Hansson, O., Minthon, L., Zetterberg, H.,
  Blennow, K., Shaw, L., Trojanowski, J.~Q., and Initiative, A. D.~N. (2010).
\newblock {Diagnosis-Independent Alzheimer Disease Biomarker Signature in
  Cognitively Normal Elderly People}.
\newblock {\em Archives of Neurology}, 67(8):949--956.

\bibitem[Morris et~al., 2010]{Morris2010}
Morris, J.~C., Roe, C.~M., Xiong, C., Fagan, A.~M., Goate, A.~M., Holtzman,
  D.~M., and Mintun, M.~A. (2010).
\newblock {APOE predicts amyloid-beta but not tau Alzheimer pathology in
  cognitively normal aging}.
\newblock {\em Annals of Neurology}, 67(1):122--131.

\bibitem[Morris et~al., 2019]{Morris2019}
Morris, J.~C., Schindler, S.~E., McCue, L.~M., Moulder, K.~L., Benzinger,
  T.~L., Cruchaga, C., Fagan, A.~M., Grant, E., Gordon, B.~A., Holtzman, D.~M.,
  and Xiong, C. (2019).
\newblock {Assessment of Racial Disparities in Biomarkers for Alzheimer
  Disease}.
\newblock {\em JAMA Neurology}, 76(3):264--273.

\bibitem[O’Hagan et~al., 2019]{OHagan2019}
O’Hagan, A., Murphy, T.~B., Scrucca, L., and Gormley, I.~C. (2019).
\newblock {Investigation of parameter uncertainty in clustering using a
  Gaussian mixture model via jackknife, bootstrap and weighted likelihood
  bootstrap}.
\newblock {\em Computational Statistics 2019 34:4}, 34(4):1779--1813.

\bibitem[Quandt and Ramsey, 1978]{Quandt1978}
Quandt, R.~E. and Ramsey, J.~B. (1978).
\newblock {Estimating mixtures of normal distributions and switching
  regressions}.
\newblock {\em Journal of the American Statistical Association},
  73(364):730--738.

\bibitem[Ruud, 1991]{Ruud1991}
Ruud, P.~A. (1991).
\newblock {Extensions of estimation methods using the EM algorithm}.
\newblock {\em Journal of Econometrics}, 49(3):305--341.

\bibitem[Sanchez et~al., 2015]{sanchez2015package}
Sanchez, L.~B., Lachos, V.~H., Moreno, E. J.~L., Sanchez, M. L.~B., and
  LazyData, T. (2015).
\newblock Package ‘censmixreg’.

\bibitem[Scrucca et~al., 2016]{Scrucca2016}
Scrucca, L., Fop, M., Murphy, T.~B., and Raftery, A.~E. (2016).
\newblock {mclust} 5: clustering, classification and density estimation using
  {G}aussian finite mixture models.
\newblock {\em The {R} Journal}, 8(1):289--317.

\bibitem[Shin and Doraiswamy, 2016]{shin2016underrepresentation}
Shin, J. and Doraiswamy, P.~M. (2016).
\newblock Underrepresentation of african-americans in alzheimer's trials: a
  call for affirmative action.
\newblock {\em Frontiers in aging neuroscience}, 8:123.

\bibitem[Vermunt and Magidson, 2021]{vermunt2021lg}
Vermunt, J.~K. and Magidson, J. (2021).
\newblock Lg-syntax user’s guide: manual for latent gold syntax module
  version 6.0.
\newblock {\em Arlington, MA: Statistical Innovations Inc}.

\bibitem[Wang et~al., 2021]{Wang2021}
Wang, W.~L., Castro, L.~M., Hsieh, W.~C., and Lin, T.~I. (2021).
\newblock {Mixtures of factor analyzers with covariates for modeling multiply
  censored dependent variables}.
\newblock {\em Statistical Papers}, 62(5):2119--2145.

\bibitem[Wang et~al., 2019]{Wang2019}
Wang, W.~L., Castro, L.~M., Lachos, V.~H., and Lin, T.~I. (2019).
\newblock {Model-based clustering of censored data via mixtures of factor
  analyzers}.
\newblock {\em Computational Statistics {\&} Data Analysis}, 140:104--121.

\bibitem[Xie et~al., 2010]{Xie2010}
Xie, B., Pan, W., and Shen, X. (2010).
\newblock {Penalized mixtures of factor analyzers with application to
  clustering high-dimensional microarray data}.
\newblock {\em Bioinformatics}, 26(4):501--508.

\bibitem[Yuksel et~al., 2012]{Yuksel2012}
Yuksel, S.~E., Wilson, J.~N., and Gader, P.~D. (2012).
\newblock {Twenty years of mixture of experts}.
\newblock {\em IEEE Transactions on Neural Networks and Learning Systems},
  23(8):1177--1193.

\bibitem[Zeller et~al., 2019]{zeller2019finite}
Zeller, C.~B., Cabral, C. R.~B., Lachos, V.~H., and Benites, L. (2019).
\newblock Finite mixture of regression models for censored data based on scale
  mixtures of normal distributions.
\newblock {\em Advances in Data Analysis and Classification}, 13(1):89--116.

\end{thebibliography}

\end{document}